\documentclass[twocolumn]{emulateapj}

\lefthead{Law et al. 2009}
\righthead{KINEMATICS OF HIGH REDSHIFT GALAXIES}

\slugcomment{DRAFT: \today}

\begin{document}

\newcommand{\msun}{\ensuremath{\rm M_\odot}}
\newcommand{\msunyr}{\ensuremath{\rm M_{\odot}\;{\rm yr}^{-1}}}
\newcommand{\Ha}{\ensuremath{\rm H\alpha}}
\newcommand{\Hb}{\ensuremath{\rm H\beta}}
\newcommand{\lya}{\ensuremath{\rm Ly\alpha}}
\newcommand{\Ntwo}{[\ion{N}{2}]}
\newcommand{\kms}{km~s\ensuremath{^{-1}\,}}
\newcommand{\ztwo}{\ensuremath{z\sim2}}
\newcommand{\zthree}{\ensuremath{z\sim3}}
\newcommand{\cgs}{\textrm{erg~s\ensuremath{^{-1}\,}~cm\ensuremath{^{-2}\,}}}
\newcommand{\cgsang}{\textrm{erg~s\ensuremath{^{-1}\,}~cm\ensuremath{^{-2}\,}~\AA\ensuremath{^{-1}\,}}}
\newcommand{\othree}{\textrm{[O\,{\sc iii}]}}
\newcommand{\otwo}{\textrm{[O\,{\sc ii}]}}
\newcommand{\oone}{\textrm{O\,{\sc i} + S\,{\sc ii}}}
\newcommand{\ntwo}{\textrm{[N\,{\sc ii}]}}
\newcommand{\ctwo}{\textrm{C\,{\sc ii}}}
\newcommand{\sitwo}{\textrm{Si\,{\sc ii}}}
\newcommand{\vsig}{\ensuremath{v_{\rm shear}/\sigma_{\rm mean}}}

\newcommand{\hst}{{\it HST}-ACS}

\title{The Kiloparsec-Scale Kinematics of High-Redshift Star-Forming Galaxies}

\author{David R.~Law\altaffilmark{1,2}, Charles~C.~Steidel\altaffilmark{3},
  Dawn~K.~Erb\altaffilmark{4}, James~E.~Larkin\altaffilmark{2}, Max~Pettini\altaffilmark{5}, Alice~E.~Shapley\altaffilmark{2},
  Shelley~A.~Wright\altaffilmark{6}}

\altaffiltext{1}{Hubble Fellow.}
\altaffiltext{2}{Department of Physics and Astronomy, University of California, Los Angeles, CA 90095;
drlaw, larkin, aes@astro.ucla.edu}
\altaffiltext{3}{Department of Astronomy, California Institute of Technology, MS 105-24,
Pasadena, CA 91125 (ccs@astro.caltech.edu)}
\altaffiltext{4}{Department of Physics, University of California, Santa Barbara, CA 93106 (dawn@physics.ucsb.edu)}
\altaffiltext{5}{Institute of Astronomy, Madingley Road, Cambridge CB3 0HA, UK (pettini@ast.cam.ac.uk)}
\altaffiltext{6}{Center for Cosmology, Department of Physics \& Astronomy, University of California, Irvine, CA 92697 (saw@uci.edu)}

\begin{abstract}

We present the results of a spectroscopic survey of the kinematic structure of star-forming galaxies at redshift $z \sim 2 - 3$ using Keck/OSIRIS
integral field spectroscopy.
Our sample is comprised of 12 galaxies between redshifts $z \sim 2.0$ and 2.5 and one galaxy at $z \sim 3.3$ which are well detected in either \Ha $\,$ or \othree $\,$ emission.
These galaxies are generally representative of the mean stellar mass of star forming galaxies at similar redshifts, although they tend to have star formation rate
surface densities slightly higher than the mean.
These observations were obtained in conjunction with the Keck laser guide star adaptive optics system, with a typical angular resolution after spatial smoothing
$\sim$ 0.15'' (approximately 1 kpc at the redshift of the target sample).
At most five of these 13 galaxies have spatially resolved velocity gradients consistent with rotation
while the remaining galaxies have relatively featureless or irregular velocity fields.
All of our galaxies show local velocity dispersions $\sim $ 60 -- 100 \kms, suggesting that (particularly for those galaxies with featureless velocity fields)
rotation about a preferred axis may not be the dominant mechanism of physical support.
While some galaxies show evidence for major mergers such evidence is unrelated to the kinematics of individual components
(one of our strongest merger candidates also exhibits unambiguous rotational structure),
refuting a simple bimodal disk/merger classification scheme.
We discuss these data in light of complementary surveys and extant UV-IR spectroscopy and photometry, concluding that the dynamical importance
of cold gas may be the primary
factor governing the observed kinematics of $z \sim 2$ galaxies.  We conclude by speculating on the importance of mechanisms for accreting
low angular-momentum gas and the early formation of quasi-spheroidal systems in the young universe.

\end{abstract}


\section{INTRODUCTION}

Galaxies are the discrete luminous building blocks of the visible universe, tracing the development of gravitational structures
across cosmic ages.  The earliest galaxies likely formed near the density peaks of the primordial
power spectrum (e.g., Bardeen et al. 1986) which were able to decouple from the cosmic expansion at early times and form a burst of stars
from the primordial gas.
As time progressed however,
these chaotic proto-galaxies gradually evolved, merging with their neighbors in newly collapsing dark matter halos,
accreting greater quantities of gaseous fuel from a filamentary intergalactic medium, and polluting their environments with the metallic detritus
of their early stellar generations.
It is at intermediate redshifts $z \sim 2 - 3$
that these morphologically irregular, juvenile galaxies
are thought to have formed the majority of the stellar
mass which we observe in modern-day galaxies
(Dickinson et al. 2003, Reddy et al. 2008).
Through a combination of galaxy--galaxy mergers, rapid star formation, and secular evolution,
these galaxies experienced a 
strong morphological transformation into the coherent structures of the Hubble sequence  which have
predominated since redshift $z \sim 1$ (Giavalisco et al. 1996; Papovich et al. 2005).

Despite our growing knowledge of the broad global characteristics of high redshift galaxies (e.g., Erb et al. 2006c; Reddy et al. 2006a;
Papovich et al. 2006; Cowie et al. 2008; and references therein), 
our understanding of their internal structure and dynamical evolution has been limited by
their small angular size, typically $\lesssim 1$ arcsecond.   
Such objects are generally not well resolved
by the ground-based imaging and spectroscopy
which form the backbone of our observational data.
This limitation precludes us from addressing such
questions as: (i) what are the triggering and
regulation mechanisms of their starbursts?
(ii) does star formation occur in the (circum)nuclear
regions of dynamically relaxed systems or as a result
of tidal shocks induced by major mergers?
(iii) do individual star-forming regions
follow a global abundance pattern or are there
strong variations in chemical enrichment within
a galaxy?
Each of these distinctions has implications
for the evolution and development of structure and stellar populations within a given galaxy.

One of the most promising means to investigate these questions is studying the ionized gas surrounding
bright star-forming regions produced by the flood of energetic photons from young stars.
Excited atoms in this gas lose energy predominantly by emission
of strong rest-frame optical nebular emission lines (redshifted into the NIR for galaxies at $z \sim 2 - 3$)
such as \Ha, \Hb, \otwo, \othree, and \ntwo.  The relative strengths of these lines encode information about the chemical composition
of the gas and the shape of the ionizing spectrum, permitting deduction of the metallicity and star formation rates
of the emission regions.  These emission features also have narrow natural line widths and are
good kinematic tracers of the ionized gas.

Such efforts using slit
spectroscopy (e.g., Pettini et al. 2001 at $z \sim 3$; Erb et al. 2004, 2006b, 2006c at $z \sim 2$; Weiner et al. 2006 at $z \sim 1$) have 
suggested that the kinematics of these galaxies may frequently be inconsistent with simple
rotationally supported gas disk models.
However, these studies have been limited by the small angular size of typical galaxies relative to the atmospheric seeing.
Often there may be only 1--2 spatially independent samples across the face of a given galaxy, and additional
uncertainty may be introduced by misalignment of the slit with the ({\it a priori}) unknown kinematic axis.  It is therefore unclear whether the high observed 
velocity dispersion
is genuine or  caused by the smearing of unresolved kinematic structure, such as disk-like rotation or merging clumps.

The recent advent of adaptive optics (AO) on 10 m class telescopes has offered
the opportunity to overcome the limitations previously imposed by atmospheric turbulence by rapidly correcting the
distorted wavefront using deformable mirrors.
Paired with integral-field spectroscopy (IFS), it is possible to obtain diagnostic spectra
of spatial regions resolved on scales of order 100 milliarcseconds (mas),
corresponding to roughly 1 kpc at redshift $z \sim 2-3$.  The data provide an empirical answer to
whether the velocity fields of galaxies in the early universe are predominantly represented by
virialized disk-like systems, major mergers, or some other dynamical structure, and whether the resulting star formation is uniform in its
properties across a given galaxy or exhibits regional variation on kiloparsec scales.

Early results from such IFS
programs with (Genzel et al. 2006; Law et al. 2007a; Wright et al. 2007, 2009)
and without (F{\"o}rster Schreiber et al. 2006; Bouch{\'e} et al. 2007; Genzel et al. 2008; Nesvadba et al. 2008; van Starkenburg et al. 2008)\footnote{One galaxy presented
by F{\"o}rster Schreiber et al. (2006) was observed with the aid of adaptive optics, as were two of the five main galaxies discussed by Genzel et al. (2008).}
adaptive optics, in addition to a small number of galaxies observed with the aid of gravitational lensing
(Nesvadba et al. 2006; Swinbank et al. 2007; Stark et al. 2008; Jones et al. {\it in prep.}), 
have helped confirm that the high velocity dispersion is an intrinsic property of high-redshift galaxies.
The total number of individual galaxies observed however is extremely small, and
considerable disagreement remains  as to the underlying nature of the dynamical structure of these galaxies, with
Genzel et al. (2008; and references therein) suggesting that rotationally supported gaseous disks form the majority of the bright galaxy population
while Law et al. (2007a) favor a significantly lower rotation fraction  with the majority of galaxies kinematically dominated by their high velocity dispersions.

In this paper we greatly expand the
original sample of Law et al. (2007a) by presenting
spatially resolved laser guide star adaptive optics (LGSAO)
spectroscopy of an additional 10
galaxies at redshifts $2.0 \lesssim z \lesssim 2.5$.
With the larger sample size it is possible to
draw broader conclusions about
the kinematical characteristics of
the galaxy population.
The paper is structured as follows.
In \S \ref{obs.sec} we describe our sample selection, observational technique, and data reduction algorithms.
In \S \ref{analysis.sec} we discuss the physical properties of our galaxies, particularly their morphological (\S \ref{gasmorph.sec}) and kinematic (\S \ref{kinematics.sec}) structure
and stellar populations (\S \ref{sedmods.sec}).
The characteristics of each of our 13 galaxies are discussed individually in detail in \S \ref{indivgals.sec}.
\S \ref{disc.sec} discusses the characteristics of the global population, comparing to recent observations by other groups in \S \ref{otherobs.sec}
and exploring possible implications for gas accretion and formation mechanisms in \S \ref{local.sec} and  \ref{theory.sec}.
We conclude with a brief summary in \S \ref{summary.sec}, while a derivation of the statistical properties of an ensemble of inclined systems is included in an
appendix (\S \ref{append.sec}).
We assume a standard $\Lambda$CDM cosmology based on 3-year WMAP data (Spergel et al. 2007) 
in which $H_0 = 73.2$ km s$^{-1}$ Mpc$^{-1}$, $\Omega_{\rm m} = 0.238$, 
and $\Omega_{\rm \Lambda} = 0.762$.

\section{OBSERVATIONS}
\label{obs.sec}

\subsection{Target Selection}
\label{tarsel.sec}

Recent years have witnessed an abundance of methods for locating galaxies in the redshift range $z \sim 2 - 3$.
These methods include optical
($U_n G {\cal R}$) color selection  (e.g., Steidel et al. 2003, 2004),
near-IR $BzK$ (Daddi et al. 2004) and $J-K$ (Franx et al. 2003) color selection, selection by sub-mm flux density
(Chapman et al. 2005), and Ly$\alpha$ emission surveys (e.g., Martin et al. 2008; and references therein).
Of these samples, those arising from the optical
color selection are perhaps the most well studied
and account for most of the global star formation
activity at $z \sim 2 - 3$ (Reddy et al. 2005).
As described in detail by Adelberger et al. (2004), the $U_n G {\cal R}$ selection technique represents a generalized version
of the Lyman-break technique employed 
by Steidel et al. (2003) to identify
rapidly star-forming galaxies at $z \sim 3$ on the
basis of their strong 912\,\AA\ Lyman break,
redshifted into the $U_n$ bandpass.
In this paper, we focus primarily on this  $U_n G {\cal R}$ sample, and particularly on those galaxies
in the redshift range $z = 1.8-2.6$ (i.e., the ``BX'' galaxy sample of Steidel et al. 2004).
While initial identification of these galaxies was based upon photometric preselection, extensive rest-UV spectroscopic follow-up work has derived
precise redshifts for all galaxies in our target sample.

The galaxies selected in this manner are typically bright and actively forming stars, with mean extinction-corrected star formation rates (SFR)
$\sim 30 \, M_{\odot}$ yr$^{-1}$ (see discussion by Erb et al. 2006b), 
and SFR surface densities similar to those observed in local starburst galaxies (e.g., Kennicutt et al. 1998b).
The resulting winds from supernovae and
massive stars drive energetic
($\sim$ a few hundred \kms; Steidel et al. {\it in prep.})
large-scale outflows into the IGM surrounding these galaxies,
creating the ubiquitous blueshifted interstellar
absorption features observed in rest-frame UV
spectra (e.g., Pettini et al. 2002; Shapley et al. 2003), and the
corresponding redshifts of the resonantly scattered
Ly$\alpha$ emission line (Verhamme et al. 2008).
With the aid of deep near-IR (e.g., Erb et al. 2006c) and mid-IR (e.g., Shapley et al. 2005b; Papovich et al. 2006; Reddy et al. 2006b; and references therein)
photometry, stellar population modeling suggests that galaxies at $z \sim 2 - 3$ span a broad range of 
stellar masses and evolutionary states.

Individual galaxies selected for study  were drawn from this pool of available targets subject to a variety of criteria.
Some targets were deliberately selected for their young stellar population ages and correspondingly small stellar masses,
some for their old ages and large stellar masses, 
and some for other reasons including
complex or multi-component rest-UV morphologies (Law et al. 2007b),
strong detections
in \Ha $\,$ narrowband surveys (Q1700-BX710 and Q1700-BX763), unusual spectral features, or previous acquisition of long-slit
kinematic data.  Given the relatively shallow  OSIRIS $K$-band sensitivity limit 
(SFR$_{\rm lim} \sim 1 \, M_{\odot}$ yr$^{-1}$ kpc$^{-2}$ in 2 hours of integration; Law et al. 2007a) 
however, the most common criterion applied
was preferential selection of those galaxies for which
previous long-slit spectroscopy (Erb et al. 2006b)
indicated nebular line fluxes
$\gtrsim 5 \times 10^{-17}$ \cgs.
We discuss the nature of our final target sample further in \S \ref{selfx.sec}.

Additional physical constraints require that there be a suitably bright ($R \lesssim 17$) star within $\sim$
60'' of each galaxy to use as a 
tip-tilt (TT) reference for the LGSAO system\footnote{The theoretical sky coverage of the Keck LGSAO system
(see, e.g., http://www2.keck.hawaii.edu/optics/lgsao/performance.html) is $\sim$ 60\%, ranging from almost 100\%
near the Galactic plane to $\sim$ 20\% or less near the Galactic poles and in particularly sparse fields such as the Hubble Deep Field.  Generally, we found
that $\sim$ 70 - 80 \% of our desired targets were close enough to a bright star to be observed with the LGSAO system.}, and that
the wavelength of redshifted
line emission falls between the strong spectroscopically unresolved night-sky OH emission features that dominate the near-IR background, and avoids wavelengths
of extremely strong telluric absorption bands.  The first of these constraints will not systematically bias the resulting sample given its reliance only on the distribution
of nearby stars, while the second imposes blackout ``windows'' in redshift space, the most significant of which spans $z \sim 2.6 - 2.9$.
The final target sample is listed in Table \ref{targets.table} and
includes 24 galaxies in 9 distinct fields distributed widely across the sky.
Of the 24 galaxies observed, 13 galaxies (distributed amongst 7 fields) have significant detections while an additional
two (Q1623-BX455 and Q2343-BX587) are too poorly detected to merit inclusion in our analysis.

\subsection{Observational Technique}
\label{obstech.sec}

Observations were performed using the OSIRIS (Larkin et al. 2006) integral-field spectrograph
in combination with the Keck II LGSAO system during 6 observing runs between June 2006 and September 2008.
The majority of these sessions followed the basic observing scheme described in \S 2.1 of Law et al. (2007a), to which we refer the reader
for a detailed description of our approach.  In brief, we observed the target galaxies in narrowband filters
(bandwidth $\sim$ 0.1 $\mu$m)
corresponding to the wavelength of either \Ha $\,$ or \othree $\, \lambda5007$ (hereafter \othree)
emission lines using the 50 mas lenslet sampling scale with 900 second individual exposures.  
Given the redshift range of most of our target galaxies ($z \sim 2 - 2.5$), \Ha $\,$ is redshifted into the $K$-band and we generally choose to observe this emission line
since the quality of the LGSAO correction is highest at longer wavelengths.  At wavelengths longer than $\sim$ 2.26 $\mu$m, however, the thermal 
continuum background
seen by OSIRIS becomes comparable to that of the atmospheric OH lines, significantly degrading the quality of the observational data.  We therefore
observe \othree $\,$ emission in either the $H$ or $K$ bands as necessary for some galaxies (e.g., Q1623-BX543 and DSF2237a-C2).
The observational configuration for each of our target galaxies is detailed in Table \ref{targets.table}.

During a typical observing sequence we first acquire a brief ($\sim 60$ second) observation of our TT reference star
in order to center our pointing and provide both a point-spread function (PSF) and flux calibration reference.  
When possible, we obtained another observation of the TT reference star at the end of our integration on a given galaxy in order to bracket the science observations
and measure any changes in the PSF.
We offset from the TT star to the target galaxy using precise offsets
measured from deep ($\sim 10$ hour) ground-based optical and/or {\it HST} Advanced Camera for Surveys (ACS) imaging data (where available).
Each galaxy was observed in pairs of 900 second exposures dithered within the field of view, typically by $\sim 1.4''$ along the long axis.
This dithering maximizes our on-source integration time but effectively halves our useful field of view to $\sim$ 1'' $\times$ 1.5'' (still significantly larger
than the typical galaxy, see discussion by Law et al. 2007b).  Each exposure pair was repeated with small $\sim 50$ mas dithers for a total of
between 1800 and 12600 seconds of integration.  
Total integration times for each object (Table \ref{targets.table}) were dictated by observational availability
and the need to obtain sufficient integration to ensure
high-quality detections.  It was typically possible to confirm detection of a given galaxy in the difference of two 900 second exposures, allowing us to
determine in real-time the targets which were likely to produce the best results given sufficient integration.
With some notable exceptions (e.g., Q2343-BX389) deeper integrations were not performed if a given galaxy was undetected within 1--2 hours.

\subsection{Data Reduction and Flux Calibration}
\label{datared.sec}

Data reduction was performed using a combination of OSIRIS pipeline and custom IDL routines described in \S 2.2 of Law et al. (2007a) which produce
a composite, three-dimensional data cube (consisting of two image dimensions and one spectral dimension) for each target galaxy.
Given the multi-year baseline over which our observations were obtained, the reduction algorithms have 
evolved over time to reflect
the changing performance characteristics of OSIRIS.
The most significant modification was introduced for data obtained in
June 2007 (i.e., for our observation of galaxies HDF-BX1564 and Q1623-BX502) which 
were affected by an imperfect focus within OSIRIS resulting in a PSF slightly elongated
along an axis roughly 45 degrees to the lenslet grid.  This was corrected by implementing a CLEAN-type algorithm (e.g., H{\"o}gbom et al. 1974) to deconvolve
the elongated PSF and replace it with a more typical circular PSF.

As described by Law et al. (2007a) the composite data cube for each galaxy is sub-sampled by a factor of two in each spatial dimension and convolved with a Gaussian
kernel in order to increase the signal-to-noise ratio (SNR) of the spectrum in each spatial pixel (spaxel).  The width of this Gaussian is chosen as required to produce
the highest quality smoothed data cube without significantly inflating the PSF delivered by the LGSAO system.  
We typically chose a FWHM of 80 mas,
although for some extremely low
surface brightness galaxies (e.g., HDF-BX1564)
we increased this smoothing to $\sim 200$ mas in order to
increase the SNR at the cost of spatial
resolution.  The effective width of the PSF (as measured from the TT star)
before and after this smoothing is given for all galaxies in Table \ref{targets.table}.  For one galaxy (HDF-BX1564)
we also smooth the spectrum of each lenslet in the wavelength domain by a kernel of FWHM 5$\,$ \AA $\,$ to better distinguish the faint emission feature from the surrounding noise.
We note that while these algorithms are largely automated, the details for each observation are carefully optimized to produce the highest-possible quality final products.

Raw composite spectra of each galaxy
(shown in Fig. \ref{spec.fig}) were obtained by
summing into a single spectrum the data from all
spaxels in a box encompassing the nebular emission
morphology, thereby sampling
both bright and faint emission regions.
This raw spectrum is calibrated using a telluric transmission spectrum
of the night sky
to normalize the throughput as a function of wavelength.  This calibration spectrum is determined
directly from observations of bright telluric standard stars for data obtained in June/September 2008, and from the OSIRIS + Keck LGSAO system + atmospheric models
of Law et al. (2006) for all previous data (see discussion in Law et al. 2007a).  The absolute flux calibration of the corrected spectra is
determined by matching extracted spectra of the TT stars to IR photometry given in the 2MASS Point-Source-Catalog.  We note that this method differs from that adopted by Law
et al. (2007a), and slight differences from their calibration are to be expected.

The largest single uncertainty in the flux calibration of these data arises from the nature of the LGSAO PSF, which can vary both with isoplanatic
angle and (rapidly) with time, resulting in significant fluctuations in the percentage of total light residing in the AO-corrected core of the PSF.  
These fluctuations are poorly understood,
and correlated primarily with mid-level atmospheric turbulence not well corrected by the AO system.
In addition low surface-brightness or extremely broad spectral features may be missed entirely (see discussion in \S \ref{selfx.sec}), 
leading to an underestimate of the total flux from galaxies with
appreciable flux in such regions.
Given the uncertainties in our bootstrapped flux calibration, we estimate that the systematic flux uncertainty for any given source is $\sim$ 30\%
(see also discussion in Law et al. 2007a).

In Table \ref{fluxes.table} we list the \othree, \Ha, and \ntwo $\,$ emission line fluxes of each galaxy (with uncertainties based on the noise
present in the underlying spectra after subtraction of Gaussian models for the identifiable emission components\footnote{Residual
features at the emission line wavelength (after subtraction of the Gaussian model profile) are generally indistinguishable from the noise in the rest of the spectrum.})
along with estimates of the oxygen abundance based on the Pettini \& Pagel (2004) calibration of the $N2 = F_{\ntwo}/F_{\Ha}$ relation.
Most of these fluxes are about half the aperture-corrected values determined from previous long-slit NIRSPEC spectroscopy (Erb et al. 2006b).
One possible explanation for this discrepancy may be that the factor of 2 aperture correction used to calibrate the NIRSPEC data could overestimate slit losses 
(particularly for the most compact galaxies), resulting in erroneously high flux estimates.  
Alternatively, this may also be due to low surface brightness emission which OSIRIS is not sufficiently sensitive to detect.

Total nebular emission line luminosity is calculated
from these fluxes assuming a standard cosmological model and correcting for extinction derived from stellar population models (\S \ref{sedmods.sec})
using a Calzetti et al. (2000) attenuation law modified as described by Erb et al. (2006c)\footnote{$E(B-V)$ ranges from 0.005 to 0.370.}.
We convert this extinction-corrected luminosity to an \Ha $\,$ star-formation rate using the Kennicutt et al. (1994) calibration:
\begin{equation}
\textrm{SFR}\;(\msunyr) = \frac{L(\Ha)}{1.26\times 10^{41} \; \textrm {erg s}^{-1}} \; \times 0.56 .
\end{equation}
where the factor of 0.56 converts to the Chabrier (2003) IMF.  For those two galaxies observed in \othree $\,$ instead of \Ha $\,$
we either assume that $\frac{L_{\Ha}}{L_{\othree}} = 1$ (Q1623-BX543, for which the actual flux ratio is unknown) or adopt
$\frac{L_{\Ha}}{L_{\othree}} = 0.52$ (DSF2237a-C2, see discussion in Law et al. 2007a).
We also list the central  wavelength of  the primary observed emission feature in each spectrum.
Systemic redshifts are determined by dividing these wavelengths by the rest-frame vacuum
wavelengths of \Ha $\,$ and \othree $\,$ (i.e., 6564.614$\,$ \AA $\,$ and 5008.239$\,$ \AA, respectively) and correcting for the heliocentric motion of the Earth
at the date, time, and
direction of observation (calculated using the IRAF {\it rvcorrect} algorithm).  We do not include fainter
\ntwo $\,$ or \othree $\,$ $\lambda4960$ in our systemic redshift calculations as these features are generally too weak to improve the redshift fit.

\begin{figure*}[tbp]
\epsscale{1.2}
\plotone{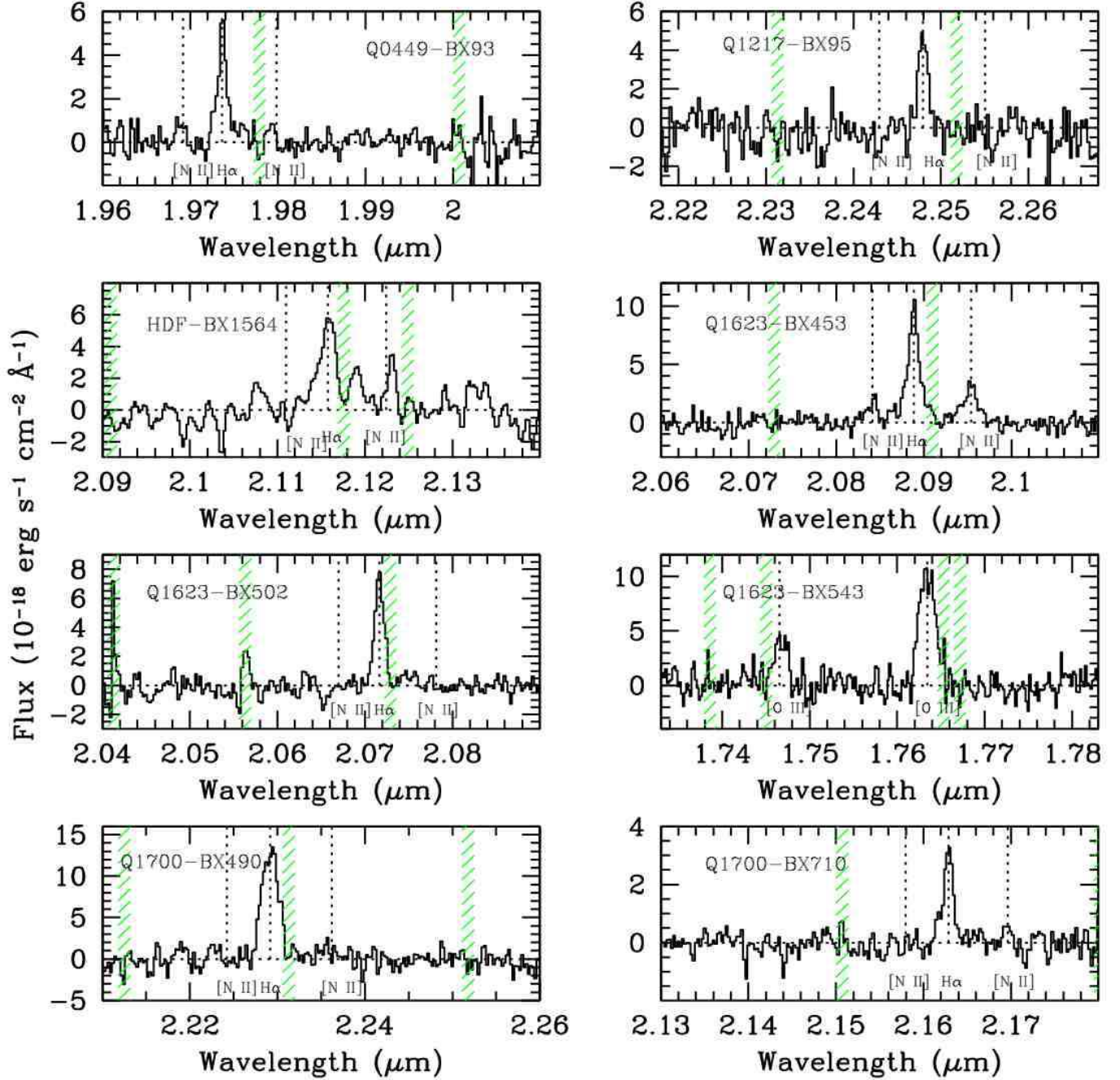}
\caption{OSIRIS spectra integrated over the spatial extent of each galaxy.  Vertical dashed lines indicate the fiducial locations
of nebular emission lines based on the systemic redshift.  Hashed green regions indicate the locations of strong atmospheric OH emission features which can give rise
to strong residuals in the spectra.}
\label{spec.fig}
\end{figure*}

\begin{figure*}[tbp]
\epsscale{1.2}
\addtocounter{figure}{-1}
\plotone{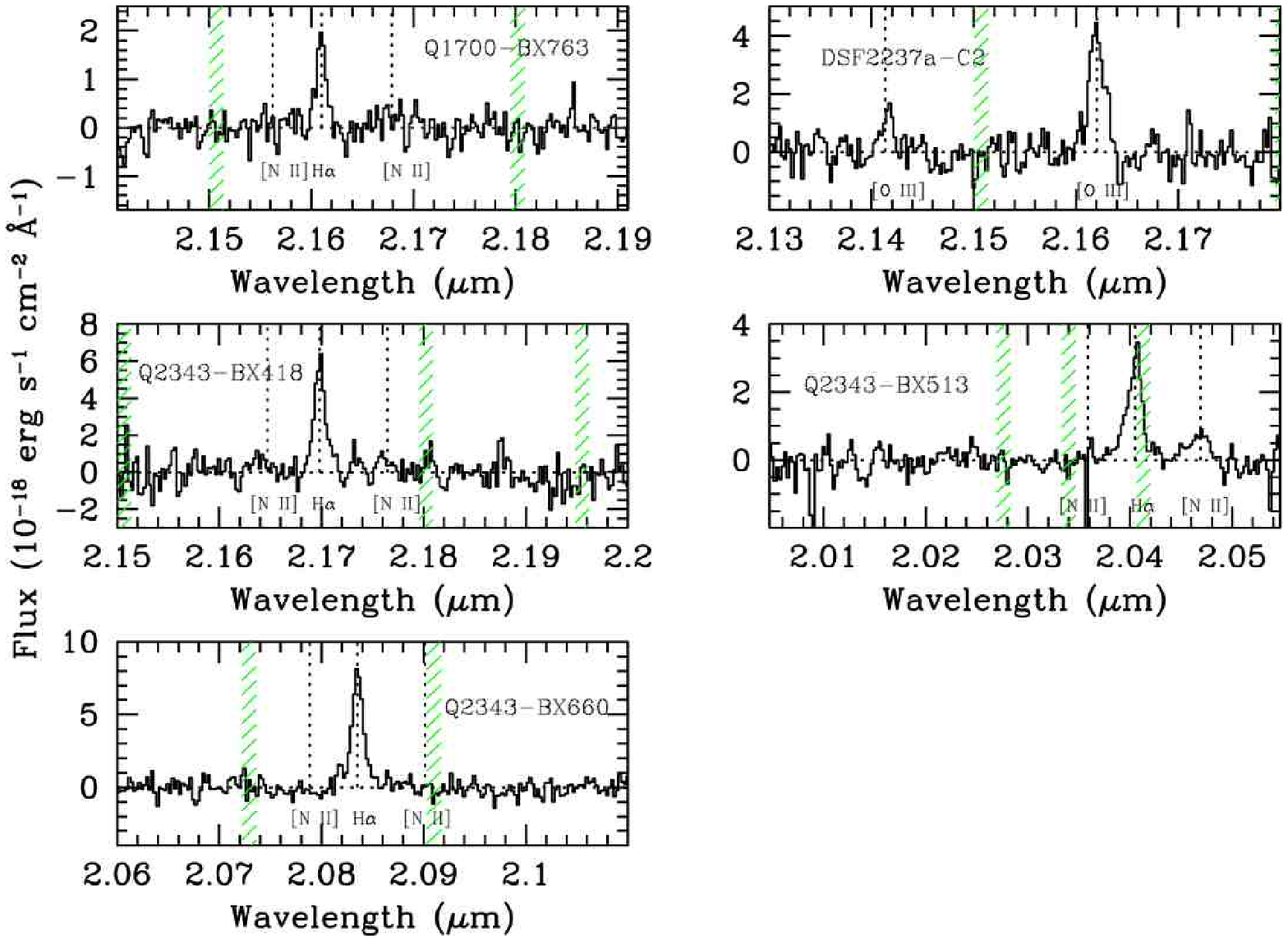}
\caption{(continued)}
\end{figure*}

Kinematic data were extracted from the oversampled composite data cubes using custom IDL routines to fit Gaussian profiles to the spectrum at each spatial location,
thereby obtaining maps of the emission line flux, wavelength centroid, and spectral FWHM across each galaxy.  This information is converted to maps
of the velocity relative to the systemic redshift and the velocity dispersion (with the instrumental resolution
[$R \sim 3600$] subtracted off in quadrature) as shown in Figure \ref{gals.fig}.
As discussed by Law et al. (2007a), the quality of these maps is improved considerably by the spatial smoothing described above, and by excluding any fits which are either unphysical or for which the data had inadequate SNR.

\begin{figure*}[tbp]
\epsscale{0.95}
\plotone{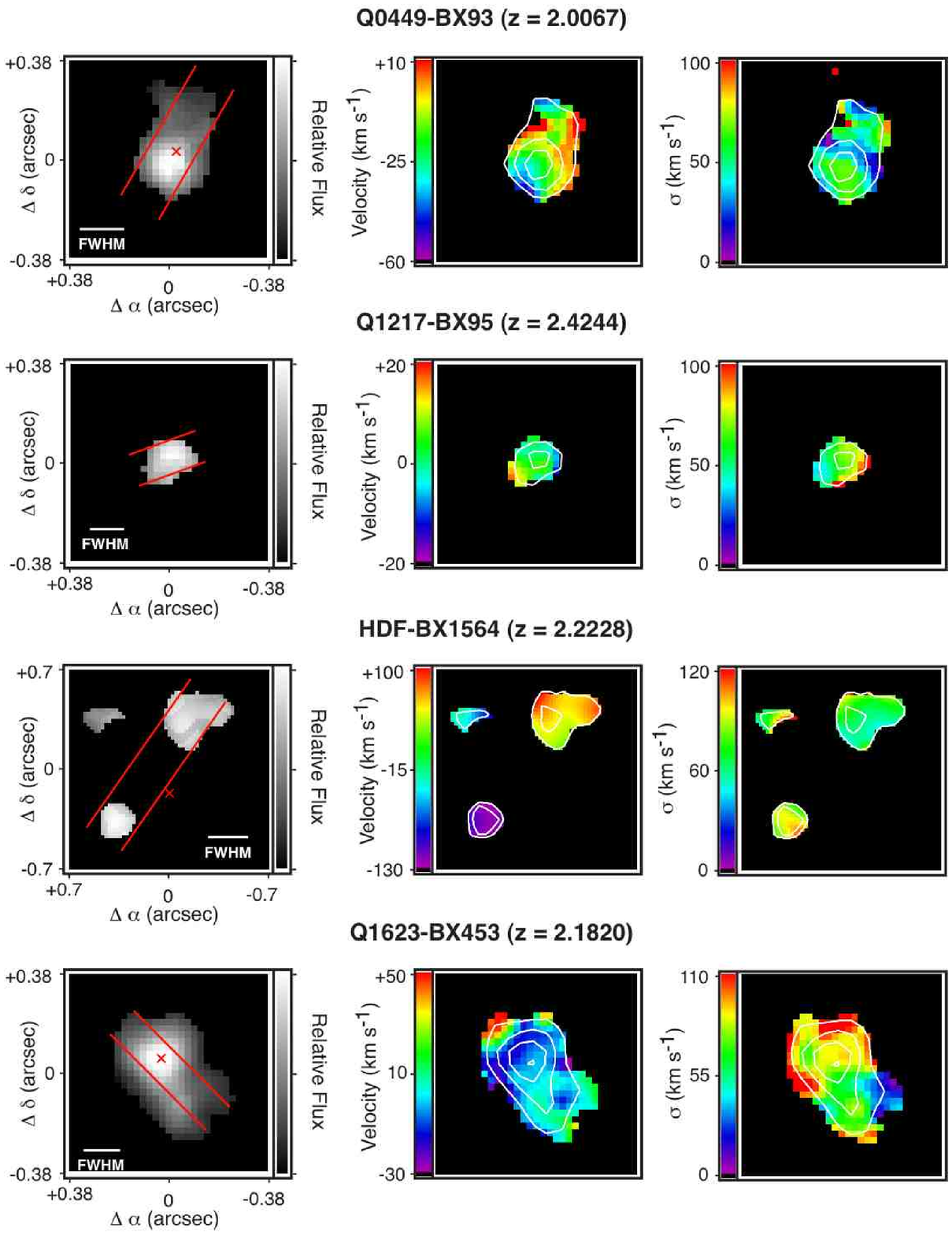}
\caption{OSIRIS maps of nebular emission (\othree $\,$ for Q1623-BX543 and DSF2237a-C2, \Ha $\,$ for all others).  Panels represent
(left to right) flux, relative velocity, and velocity dispersion maps.  Individual pixels measure 25 mas, the total field of view
varies from target to target as needed to contain the emission line regions.  The FWHM of the PSF (after smoothing described in \S \ref{datared.sec}) is indicated 
in the left-hand panel for each galaxy.
Contours represent linear intervals in line flux density.  All images are presented in a standard orientation with North up, and East left.
Red lines indicate the `slits' used to extract the 1-d velocity curves shown in Figure \ref{slitcurves.fig}.
Red X marks indicate the location of peak \ntwo $\,$ emission (when present).  Lone pixels are likely noise artifacts rather than genuine galaxy features.}
\label{gals.fig}
\end{figure*}

\begin{figure*}[tbp]
\addtocounter{figure}{-1}
\plotone{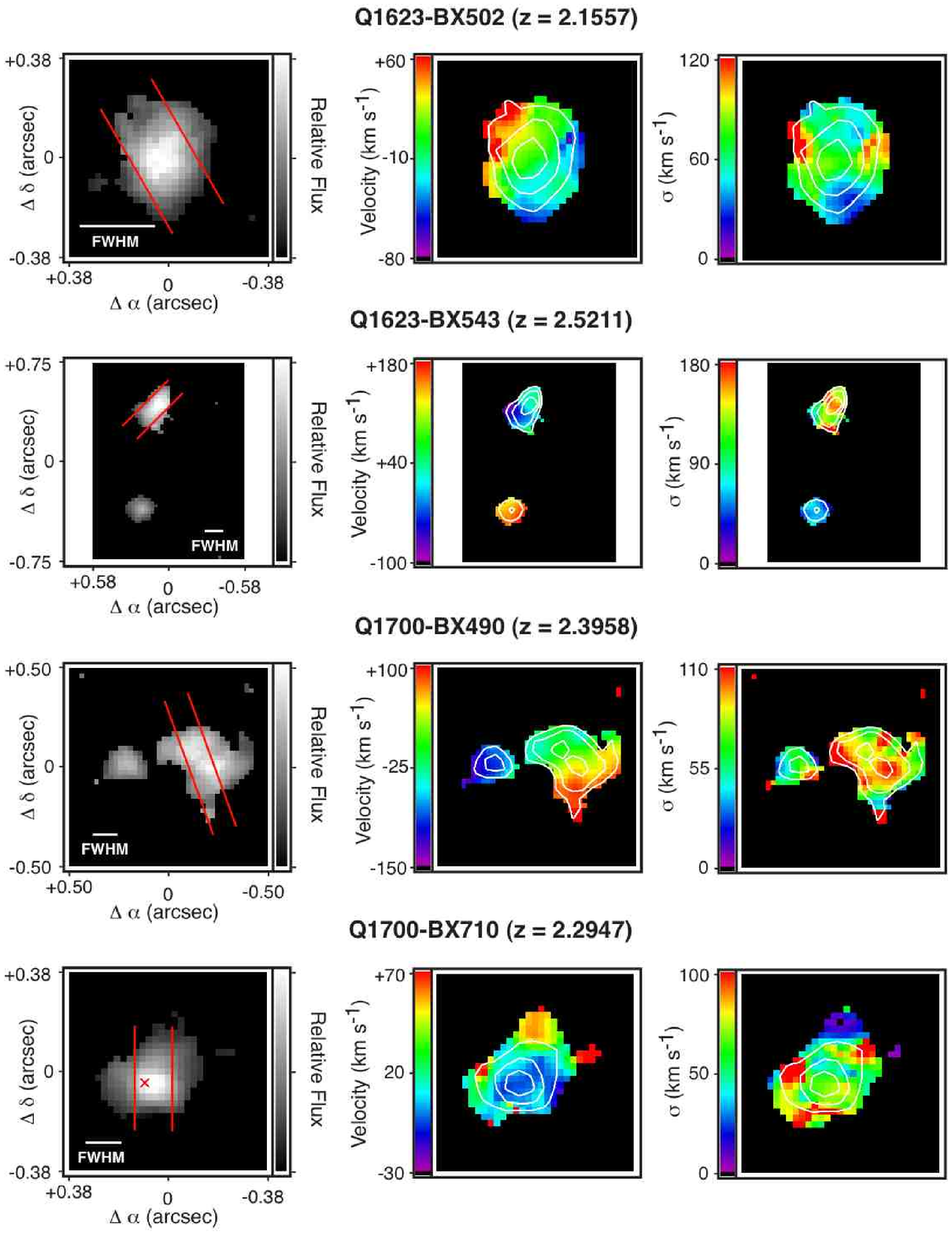}
\caption{(continued)}
\end{figure*}

\begin{figure*}[tbp]
\addtocounter{figure}{-1}
\plotone{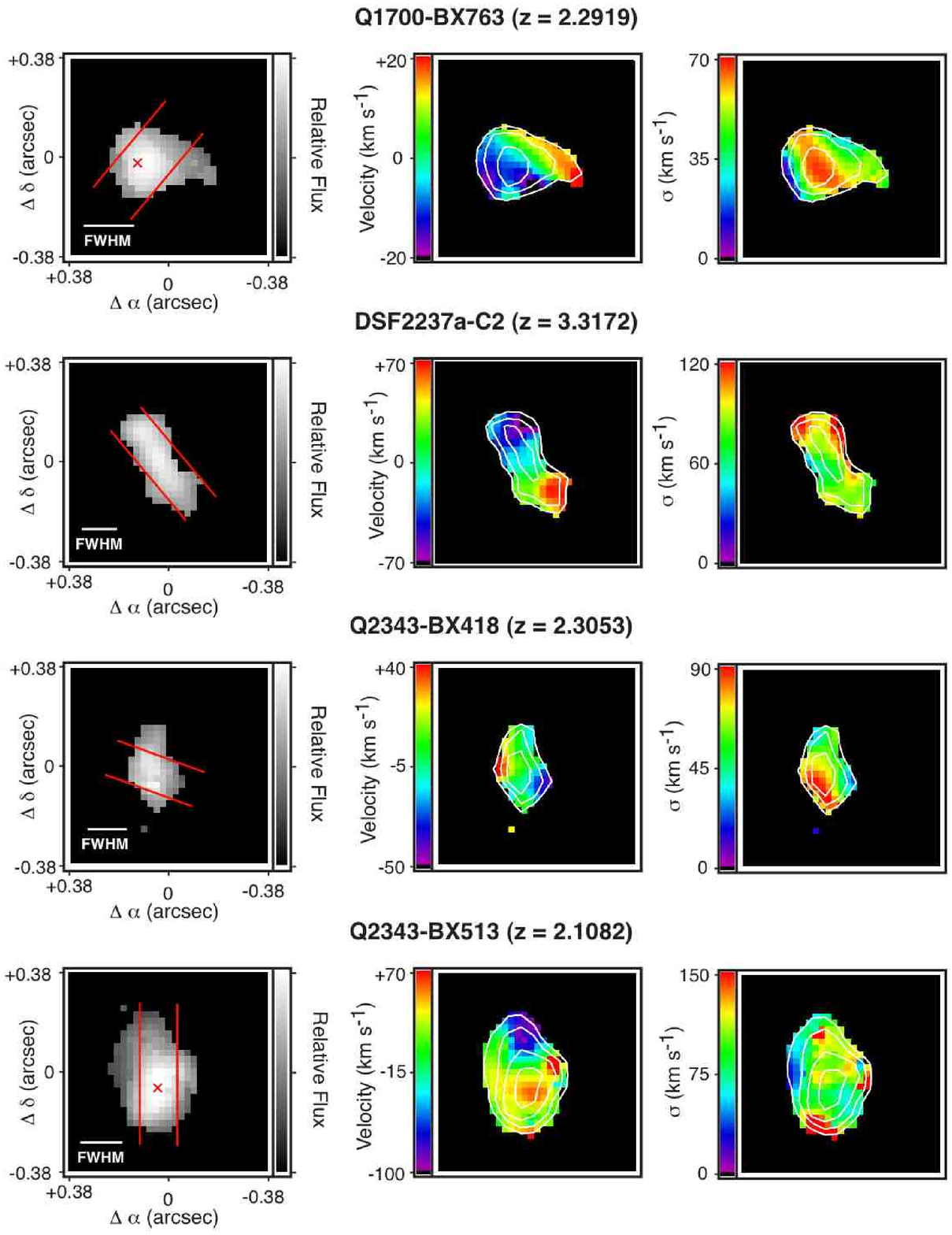}
\caption{(continued)}
\end{figure*}

\begin{figure*}[tbp]
\addtocounter{figure}{-1}
\plotone{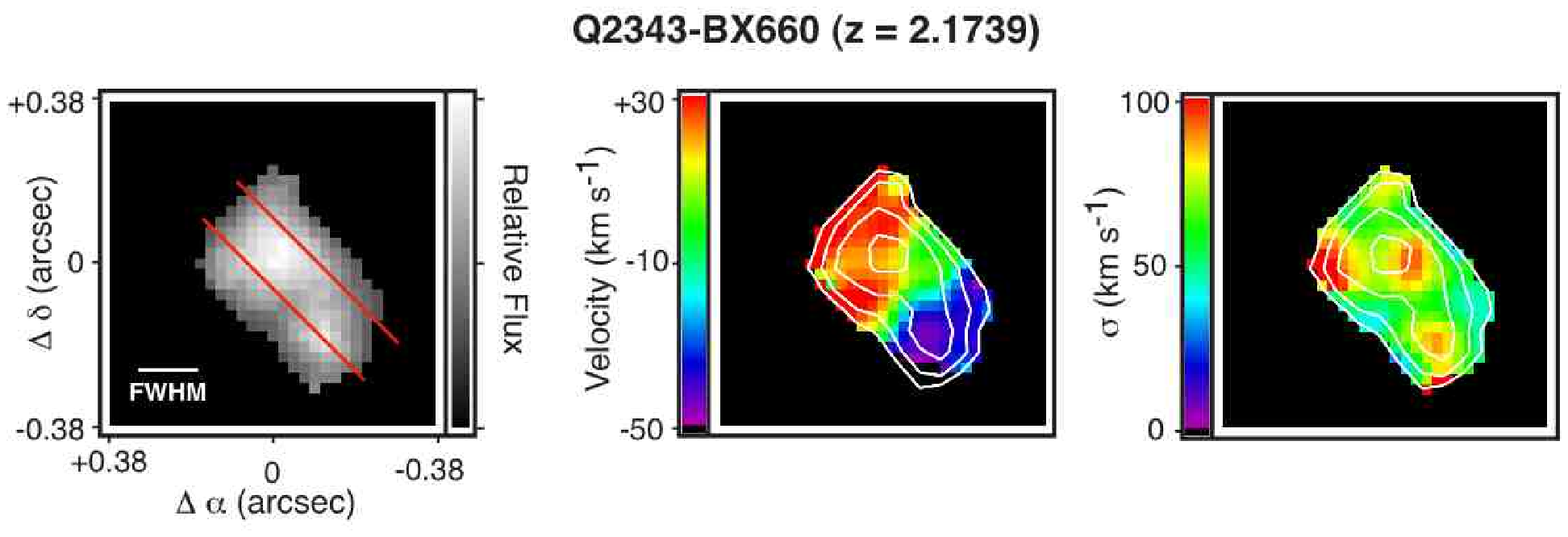}
\caption{(continued)}
\end{figure*}

\section{Results}
\label{analysis.sec}
\subsection{Morphologies}
\label{gasmorph.sec}

As demonstrated by numerous authors (e.g. Abraham et al. 1996; Conselice et al. 2005; Elmegreen et al. 2005; Lotz et al. 2006; Law et al. 2007b) the morphological structure of galaxies
at redshift $z \gtrsim 2$ is highly irregular and frequently composed of multiple spatially separated components when observed in the rest-frame UV.  It is worthwhile exploring the similarity
between such rest-UV morphologies and those of the ionized gas.
Given the extremely faint nature of these galaxies, we produce maps of the ionized gas morphology (Fig. \ref{gals.fig}, left-hand panels) by summing spectral channels within 1$\sigma$ of the peak emission wavelength at each
oversampled spatial pixel meeting the SNR criteria described in \S \ref{kinematics.sec}.  Such a `tunable' narrow-band filter for each position in the galaxy minimizes the contribution of noise
to the flux map .

Following the method outlined by Abraham et al. (2007) we define a segmentation map for each galaxy using a quasi-Petrosian isophotal cut ($\eta = 0.2$) which is independent of
morphology.  Using this segmentation map we calculate a variety of morphological statistics including the Gini coefficient $G$ (a measure of the curve of growth of the light distribution, see discussion
by Abraham et al. 2003), the multiplicity parameter $\Psi$ (a measure of the number of components of the light distribution, see Law et al. 2007b), the second-order moment $M_{20}$ of those pixels constituting
the top 20\% of the total flux (Lotz et al. 2004), the total luminous area $I$ and effective radius $r$ corrected for the PSF as described in Law et al. (2007a), and the distance $d_{\rm 2c}$ between individual
components (for galaxies which have spatially separated pieces).  These measurements are given in Table \ref{morphs.table}.

We note, however, that it is difficult to compare these values directly to rest-UV morphological data presented by Lotz et al. (2004, 2006) and Law et al. (2007b) due to systematic differences in the
observational data.
First (and most important) the quasi-Petrosian isophotal
cut simply selects all pixels above our S/N ratio threshold
for nearly all of our galaxies since our method of constructing the lowest-noise flux maps has 
artificially eliminated all of the fainter galaxy pixels for which it was not possible to fit a reliable emission-line spectrum.  
While this is clearly not ideal from a morphological standpoint, 
for our lowest surface-brightness sources (e.g. HDF-BX1564) this spectral line-fitting method and associated SNR cut is frequently the only mechanism by which we can reliably distinguish
any features from the background noise at all.
Our isophotal cut generally does not reach
surface brightness levels as low as those of many narrow-band
studies; this results in lower values
of $G$ than are normally derived
in the rest-frame UV (e.g., Lotz et al. 2004, 2006; Law et al. 2007b) as there is less contrast among pixels in the segmentation map.
In contrast, the multiplicity parameter $\Psi$ is relatively robust to such surface brightness variations and still reliably indicates the presence of multiple components
in the light profile.
Additionally, many of the galaxies observed are only a few times the size of the observational PSF (which can vary considerably from galaxy to galaxy) 
which also has an impact on the numerical classification
of the morphology.
We therefore recommend that these morphological statistics
be interpreted in a relative, rather than absolute, sense.

\begin{figure*}
\epsscale{0.7}
\plotone{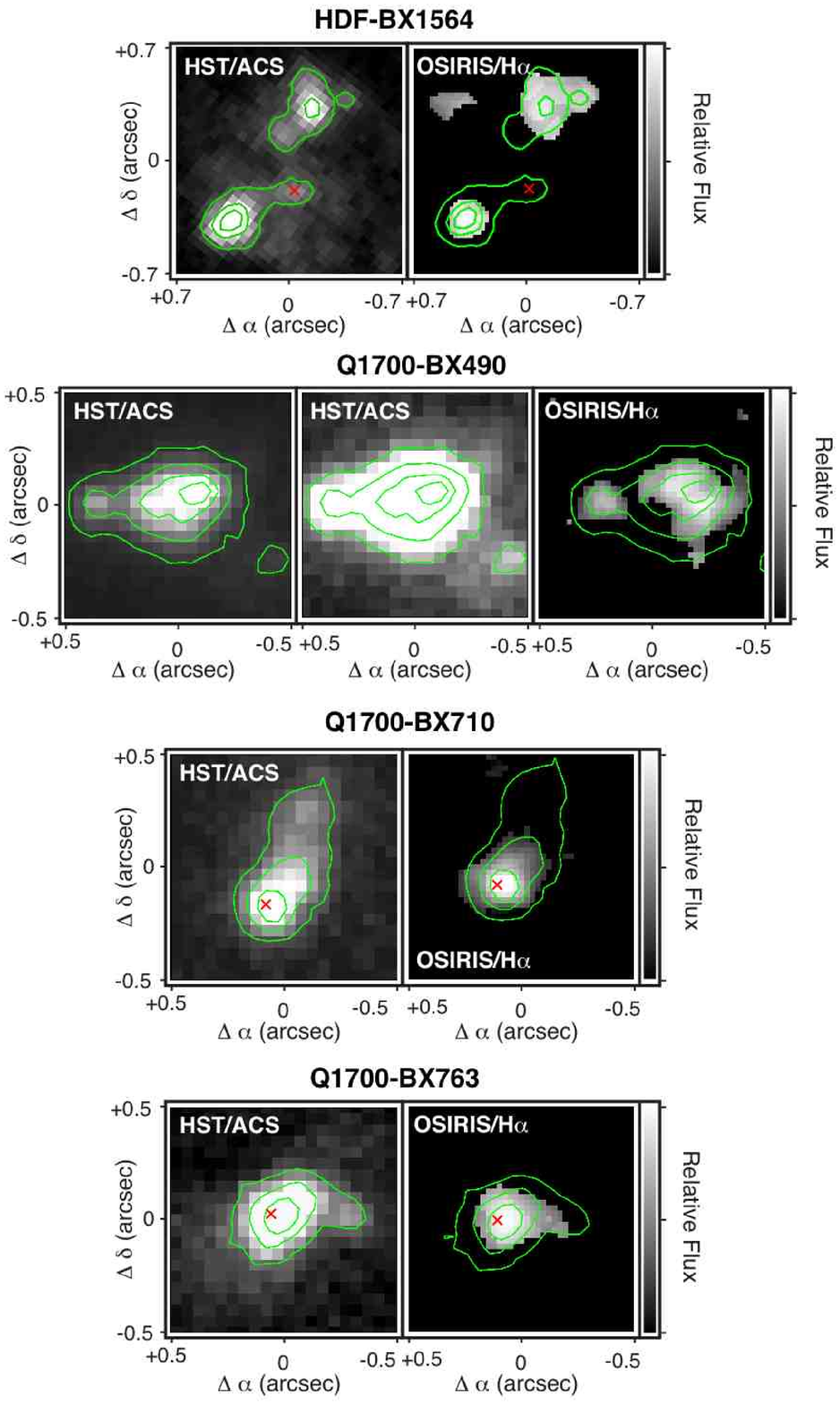}
\caption{Comparison between \hst\ rest-UV continuum morphologies and OSIRIS nebular emission morphologies.  
In each panel the grayscale and contours are linear
in rest-UV flux density.  Two copies of the \hst\ map for Q1700-BX490 are included with different greyscale stretches to best emphasize both
bright and extended faint structure.  Note that the fields of view for Q1700-BX710 and Q1700-BX763 are slightly larger than those in Figure \ref{gals.fig}
in order to completely encompass the extended \hst\ morphology.  Red X marks indicate the centroid of peak \ntwo $\,$ emission (when present).}
\label{hst.fig}
\end{figure*}

Four of our target galaxies (HDF-BX1564, Q1700-BX490, Q1700-BX710, and Q1700-BX763) lie in fields for which deep optical imaging data has been
obtained with \hst.  These data in the HDF and Q1700 fields have been described by Law et al. (2007b) and Peter et al. (2007), respectively.
The comparative morphologies of \hst $\,$ (i.e., tracing the rest-UV continuum) and OSIRIS \Ha $\,$ emission are shown in Figure \ref{hst.fig}.
Given the factor of $\sim 5$ difference between the limiting SFR surface density probed by the \hst\ and OSIRIS data ($\sim 0.2 M_{\odot}$ yr$^{-1}$ kpc$^{-2}$ and
$\sim 1 M_{\odot}$ yr$^{-1}$ kpc$^{-2}$ respectively; see discussion in Law et al. 2007a) there
is generally a good correspondence between the respective morphologies.
Due to the extremely narrow field of view of OSIRIS, there are no absolute reference points to which we can calibrate the coordinate solution as
our offsets from the TT acquisition star are only reliable to within $\sim 100$ mas (i.e., comparable to our PSF)
due to various uncertainties resulting from (for example) measuring the coordinates of the TT star, global uncertainties in the coordinate solution of the ground based imaging data,
and the unknown proper motion of the TT stars.
It is not possible therefore to reliably align the \hst $\,$ and OSIRIS images to high precision using absolute coordinates alone.
Instead, we align the \hst $\,$ contours shown in Figure \ref{hst.fig} 
with the OSIRIS maps ``by-eye'', sliding the contours freely until they appear to best overlap the \Ha $\,$ flux data.
While this prohibits us from investigating small discrepancies between the locations of peak emission in isolated single sources, we are still able to compare
spatial differences for sources with multiple emission features and the relative shapes of the various emission profiles.
We discuss the specific features of individual galaxies in \S \ref{q0449bx93.sec} -- \ref{q2343bx660.sec} below.

\subsection{Stellar Population Modelling and Gas Masses}
\label{sedmods.sec}

Using our extensive ground-based $U_n G {\cal R} J K_s$ and {\it Spitzer} IRAC and MIPS photometry in the selected
survey fields it is possible to construct stellar population models for the spectral energy distributions  of the target galaxies.
The modeling procedure is described in detail by Shapley et al. (2005a).  In brief, we use Bruzual and Charlot (2003) models
with a constant star formation rate (unless otherwise specified), solar metallicity, and a Chabrier (2003) initial mass function (IMF).
These models are overplotted on our photometric data in Figure \ref{SEDplots.fig} and the individual values for
SFR, stellar mass ($M_{\ast}$), population age, and extinction are tabulated in Table \ref{SED.table}.
As discussed by previous authors (e.g., Shapley et al. 2001, 2005a; Papovich et al. 2001) ,
degeneracies in the stellar population model tend to cancel out with regard to the total stellar mass, meaning that this is typically the
best constrained of these parameters.
We note that previous modeling of many of these galaxies has been presented by Erb et al. (2006c) and Shapley et al. (2001);
the previous models are updated here for consistency and to add {\it Spitzer} IRAC data where available.

\begin{figure*}[tbp]
\epsscale{1.0}
\plotone{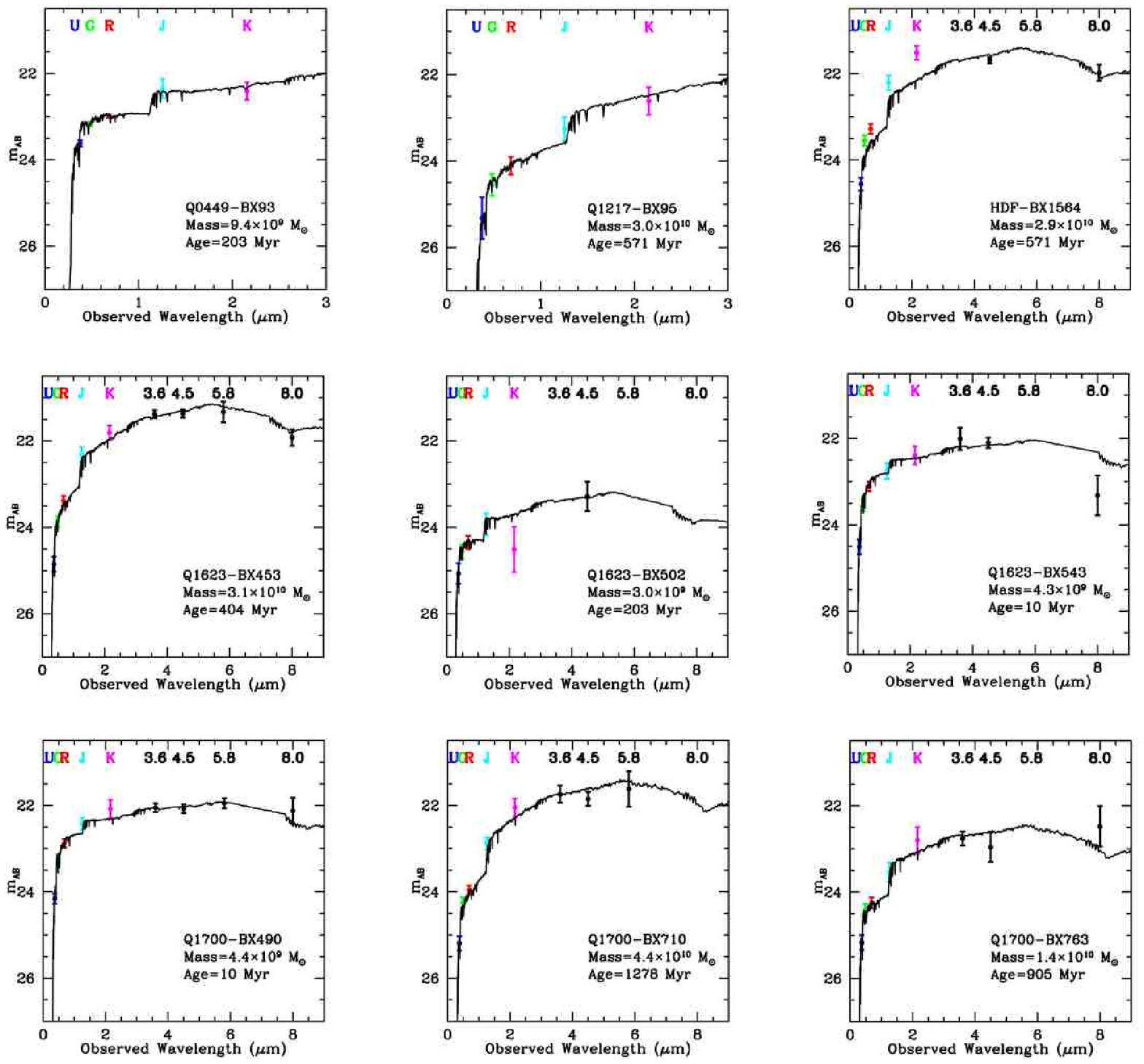}
\caption{The best-fit constant star formation (CSF) model (solid black line) is overplotted against the observed spectral energy distribution for
our target galaxies. Colored points represent ground-based optical and near-IR photometry; black points are based on \textit{Spitzer}-IRAC observations.
Values given for stellar mass and population age represent the values derived from the best-fit CSF model using a Chabrier (2003) IMF; typical uncertainties are given
in Table \ref{SED.table}.}
\label{SEDplots.fig}
\end{figure*}

\begin{figure*}[tbp]
\addtocounter{figure}{-1}
\epsscale{1.0}
\plotone{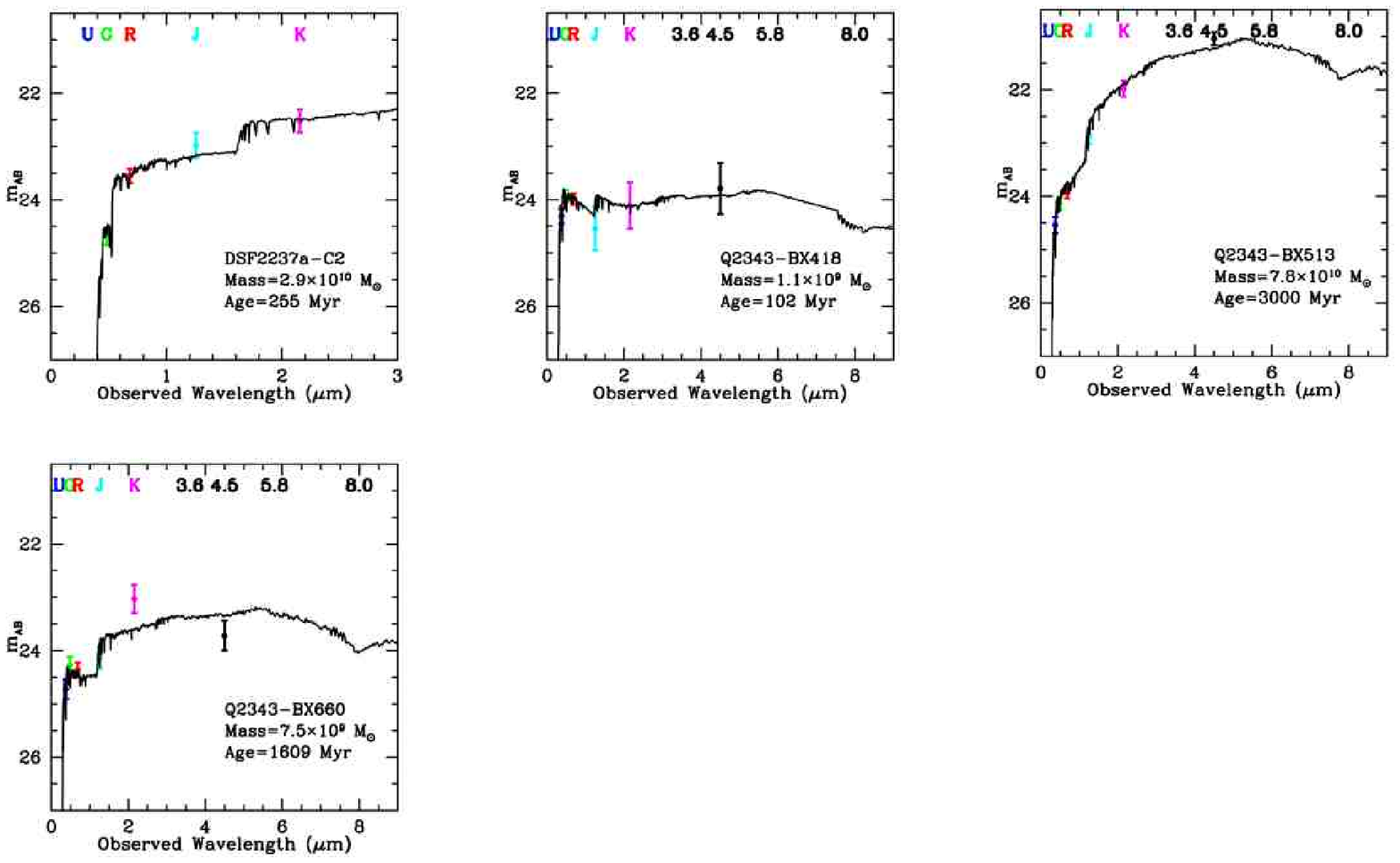}
\caption{(continued)}
\end{figure*}

The galaxies for which the poorest fit to the photometry is obtained 
(HDF-BX1564, Q1623-BX502, Q1623-BX543, Q2343-BX660)
tend to be those with multiple components to the light distribution.
In general, however, there is reasonable agreement
between the values of SFR derived from SED fitting
and those implied by the \Ha\ luminosity (Table \ref{SED.table}),
although the SED-based estimates tend to be
systematically greater than those based on \Ha.

We also include in Table \ref{SED.table} estimates of the mass in cold gas of each of our galaxies (and the corresponding gas fraction $\mu = M_{\rm gas} / (M_{\rm gas} + M_{\ast}$))
determined by Erb et al. (2006c) based on long-slit NIRSPEC spectroscopy.  
While it would be possible to use the global Schmidt law (Kennicutt 1998) to calculate new estimates of $M_{\rm gas}$ from our observed \Ha\
fluxes, the results of this calculation are sensitive to both \Ha\ flux and the apparent size of the galaxy, which systematically differ between OSIRIS and NIRSPEC due to the different 
surface brightness threshholds reached by the two spectrographs.  We use the Erb et al. (2006c) estimates for consistency in our later (\S \ref{otherobs.sec}) comparison
to other galaxy samples.

\subsection{Kinematics}
\label{kinematics.sec}

As illustrated by Figure \ref{gals.fig} (middle panels), galaxies show a mix of velocity fields ranging from largely featureless (Q1623-BX453)
to smoothly varying (Q1700-BX490) to extremely complex (Q2343-BX513).  
In particular, we find that five of our galaxies (Q1623-BX502, Q1623-BX543, Q1700-BX490, DFS2237a-C2, Q2343-BX418)
have velocity fields consistent with rotation-like velocity gradients, six (Q0449-BX93, Q1217-BX95, HDF-BX1564, Q1623-BX453, Q1700-BX710, Q1700-BX763) consistent with negligible
velocity structure, and two (Q2343-BX513, Q2343-BX660) with velocity structure unlike simple rotation.  
Additionally, three of our 13 galaxies (HDF-BX1564, Q1623-BX543, Q1700-BX490) consist of multiple spatially (and kinematically) distinct
regions: two of these galaxies display significant kinematic shear in their individual components (Q1623-BX543, Q1700-BX490) while
one does not (HDF-BX1564).

We discuss the kinematics of individual galaxies in detail in \S \ref{indivgals.sec} below.
The single feature common to all of these galaxies however is the significant velocity component with no preferred kinematic axis
(Fig. \ref{gals.fig}, right-hand panels) which dominates over all coherent kinematic structures.
This velocity dispersion may be quantified in two ways.  An estimate of the overall velocity dispersion $\sigma_{\rm net}$ of  a galaxy may be obtained by simply
fitting a single Gaussian profile to the spatially integrated spectrum of the entire galaxy (i.e., the spectra shown in Fig. \ref{spec.fig}).  
This measurement does not distinguish between spatially resolved
velocity gradients and small-scale motions within a given resolution element.
While $\sigma_{\rm net}$ can provide a reasonable estimate of the dynamical mass using the formula
\begin{equation}
M_{\rm dyn} = \frac{C \sigma_{\rm net}^2 r}{G}
\label{dynmass.eqn}
\end{equation}
(where $C = 5$ for a uniform sphere; see Erb et al. 2006c), it is not an optimal means of determining the intrinsic dispersion of the ionized gas.
We therefore also calculate $\sigma_{\rm mean}$ which is the flux-weighted mean\footnote{We adopt a flux -weighted mean rather
than a simple mean in order to minimize contamination from the lowest S/N ratio spectra.}
 of the velocity dispersions measured in each individual spaxel 
(i.e., the `local' velocity dispersion shown in the right-hand panels of Figure \ref{gals.fig}).
This quantity effectively suppresses spatially resolved velocity gradients
and provides a more accurate measure of the typical line of sight velocity dispersion at a given location within the galaxy, albeit with some potential bias due to the flux weighting,
beam smearing, and finite sampling of the observational data.
Values of $\sigma_{\rm mean}$ and $\sigma_{\rm net}$
are tabulated in Table \ref{kinematics.table}.
In the three cases where the galaxy consists of spatially
distinct regions
we calculate values for $\sigma_{\rm mean}$ in each of the regions
and also measure the kinematic offset $v_{\rm 2c}$ between them.
As expected, values of $\sigma_{\rm mean}$ are generally consistent with or slightly less than $\sigma_{\rm net}$
except for cases where large-scale velocity structure inflates $\sigma_{\rm net}$ (e.g. HDF-BX1564).

\begin{figure*}[tbp]
\plotone{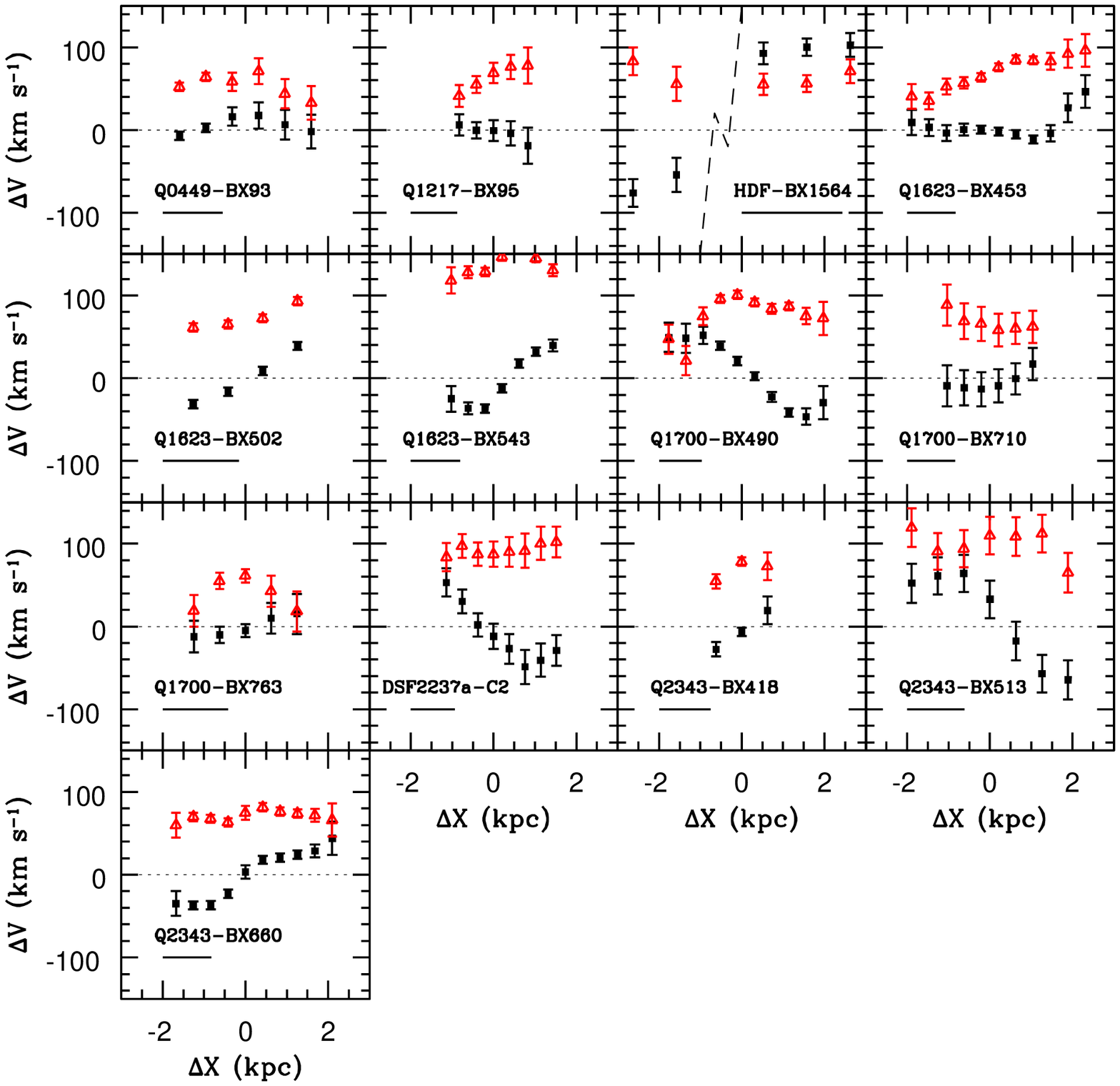}
\caption{Relative velocity (black filled squares) and velocity dispersion (red open triangles) curves along the kinematic major axis for each galaxy (see Fig. \ref{gals.fig} for orientation).
The solid line in each panel represents the FWHM of each observation, the sampling scale in each panel corresponds to half of this value to minimize spatial correlation
between individual points.  Error bars represent uncertainties derived from numerical Monte-Carlo modelling.  Note that the plot
for HDF-BX1564 is discontinuous and omits a 4 kpc region of blank sky between the two components in order to fit both into the same display area as the other galaxies.}
\label{slitcurves.fig}
\end{figure*}

We place these descriptive arguments on a firmer numerical footing by constructing one-dimensional velocity curves of each of our target galaxies 
as shown in Figure \ref{slitcurves.fig}.  
These are obtained by extracting spectra
from our calibrated data cubes along simulated slits
whose width is matched to the spatial resolution of the data, 
 and orientations 
(indicated by red lines in Fig. \ref{gals.fig}) chosen to maximize the apparent velocity gradient across the galaxy.
We define the shear velocity $v_{\rm shear}$ as half the maximum difference between any two positions along these pseudo-slits, noting
that this will disregard whether the maximum difference occurs between locations at the extreme ends of a galaxy or indeed whether the resulting velocity
curves resembles actual rotation in any way.  Any apparent trends can easily be confirmed by inspection of the two-dimensional velocity map (Fig. \ref{gals.fig}).

Ideally, it would be possible to perform detailed kinemetric analyses of each of our galaxies 
(e.g. Shapiro et al. 2008; Genzel et al. 2008; Wright et al. 2009) in order to obtain a set of
best-fitting models for the light distribution and ionized gas kinematics, along with the dynamical ratio
$v_{\rm disk}/\sigma_{\rm int}$ of the model (where $v_{\rm disk}$ is the maximum circular speed
attained within the galaxy and $\sigma_{\rm int}$ is the intrinsic local velocity dispersion).
However, the observational data generally do not justify such detailed analyses because the majority of the galaxies do not show
concrete evidence (such as spatially resolved velocity shear) to suggest that they might be
described by traditional inclined-disk models.
Of the 5 galaxies with smoothly
varying velocity gradients consistent with rotation, only two have these gradients
aligned with the morphological major axis
(as might often be expected for a foreshortened disk).
Of these two, one (Q1623-BX543) has a close ($\sim 7$ kpc)
companion which does not partake in the same pattern
of velocity shear.
A model for the single remaining galaxy, DSF2237a-C2,
which is isolated and has a significant kinematic
gradient aligned with its morphological major axis
has been presented by Law et al. (2007a).

Absent such models, the best kinematic statistic available is the observed ratio \vsig, values of which range 
from $0.2 - 0.7$ with a typical uncertainty $\sim 0.2$ (Table \ref{kinematics.table}).
While we have attempted to maximize the value of \vsig by appropriate choice of the position angle for our pseudo-slit, and by using the smaller of our two estimates of
$\sigma$, the meaning of any particular value must be considered with regard to several important factors.
\begin{enumerate}
\item The observed velocity curves will be foreshortened by a factor of sin $i$.  On average this will introduce a correction factor of $\pi$/4 (see \ref{append.sec}).
\item The OSIRIS data are relatively shallow and probe radii $\lesssim 2$ kpc.  If the velocity curve rises at larger radii, our 
observations will underestimate the true circular velocity of the system.
\item Unresolved kinematic structure smeared by the observational beam will decrease the observed velocity gradient and 
inflate the central velocity dispersion to which the flux-weighted $\sigma_{\rm mean}$ is most sensitive.
In the case of a rotating disk model this effect will
be particularly pronounced at small radii where the velocity gradient is steepest.
\end{enumerate}

All of these effects will systematically produce values of \vsig lower than $v_{\rm disk}/\sigma_{\rm int}$, although
the magnitude of these effects  is difficult to quantify as they depend strongly upon the (unknown) intrinsic kinematic structure of the galaxies.
Rotating disk models may expect correction factors $\sim 2 - 4$ (the disk model for DSF2237a-C2 developed by Law et al. 2007a, for example, had a correction
factor $\sim 2$), depending on the size of the observational PSF and the precise distribution of flux with respect to the underlying rotation field.
Even allowing for an extreme correction factor $\sim 5$, the observed range $\vsig = 0.2 - 0.7$ is significantly lower than the typical values  ($v_{\rm disk}/\sigma_{\rm int} \sim 10-20$, i.e.,
\vsig $\sim 2-4$) appropriate to disk galaxies today (e.g. Dib et al. 2006), in agreement with the general conclusion 
that these galaxies must be considerably thicker than traditional disk galaxies in the local universe
(see also discussion by F{\"o}rster Schreiber et al. 2006, Genzel et al. 2006, Law et al. 2007a, Genzel et al. 2008).

Such corrections to the observational data may  not obviously be desirable however for galaxies which are not
well-described by inclined disk models.  
In at least six of our 13 galaxies both the velocity and velocity dispersion are consistent with a constant value across multiple resolution elements.
In the absence of coherent velocity structure, the  high velocity dispersions observed for $z \sim 2$ galaxies ($\langle \sigma_{\rm mean}\rangle = 78\pm17$ \kms) cannot be explained
by the convolution of such structure with the observational PSF\footnote{Numerical modelling suggests  
that beam-smearing can account for at most a few percent of the observed $\langle \sigma_{\rm mean}\rangle$.}
(as, for instance, in the $z \sim 1.6$ sample of Wright et al. 2009), and instead indicate that both
the high velocity dispersion and low bulk motions about a preferred kinematic axis may be intrinsic properties of the ionized gas.

\section{Notes on Individual Galaxies}
\label{indivgals.sec}

\subsection{Q0449-BX93}
\label{q0449bx93.sec}

This intermediate-mass galaxy ($M_{\ast} = 0.9 \times 10^{10} M_{\odot}$) is dominated by a single emission component at the systemic redshift with a faint
secondary component offset by $\sim$ 150 \kms located $\sim$ 1 kpc to the northeast.  
This faint secondary feature may represent a kinematically distinct star-forming region or a small galaxy in the process of merging
with the the brighter system (see discussion by Law et al. 2007a).
Weak \ntwo $\,$ emission (both $\lambda 6549$ and $\lambda 6585$
lines are detected with similar significance) is roughly concentric
with the \Ha $\,$ emission and implies a metallicity
$12 + \log({\rm O/H}) = 8.37\pm0.09$ (approximately half-solar).
There is no evidence for significant bulk motions with $\vsig = 0.2 \pm 0.2$.

\subsection{Q1217-BX95}

This galaxy is qualitatively similar to Q0449-BX93, and is well-detected but compact (only slightly larger than the observational PSF)
with no evidence of extended structure despite its moderate stellar mass ($3 \times 10^{10} M_{\odot}$).
There is no obvious velocity structure within this source.

\subsection{HDF-BX1564}

HDF-BX1564 is among the least well detected of our galaxies, and required heavy spatial and spectral smoothing of the composite data cube
in order to successfully discern emission features.  This extensive processing has the unfortunate consequence of increasing our effective PSF
to $\sim$ 0.3'', comparable in size to the two primary emission regions (northwest [NW] and southeast [SE]) comprising the source.  
These two features are separated by $v_{\rm 2c} \sim$ 171 \kms 
and $d_{\rm 2c} = 7$ kpc.  
While a third component is apparent in Figure \ref{gals.fig}, based on analysis of typical noise patterns within the composite data cube we believe that this feature is 
probably not real.
Supporting this explanation is the rest-UV morphology, which we show for comparison in Figure \ref{hst.fig}.  Both the NW and SE features
have counterparts in the ACS imaging data, while the questionable northeastern feature does not.  The \hst $\,$ imaging data also suggests 
faint diffuse emission in the field of view, particularly in the form of spurs reaching between the bright knots.  Given that the bright knots themselves are only
barely detected (at a S/N ratio of $\sim$ 5) in the heavily smoothed OSIRIS data, it is not surprising that we do not detect these spurs, even if they have \Ha $\,$ counterparts.

\begin{figure}
\plotone{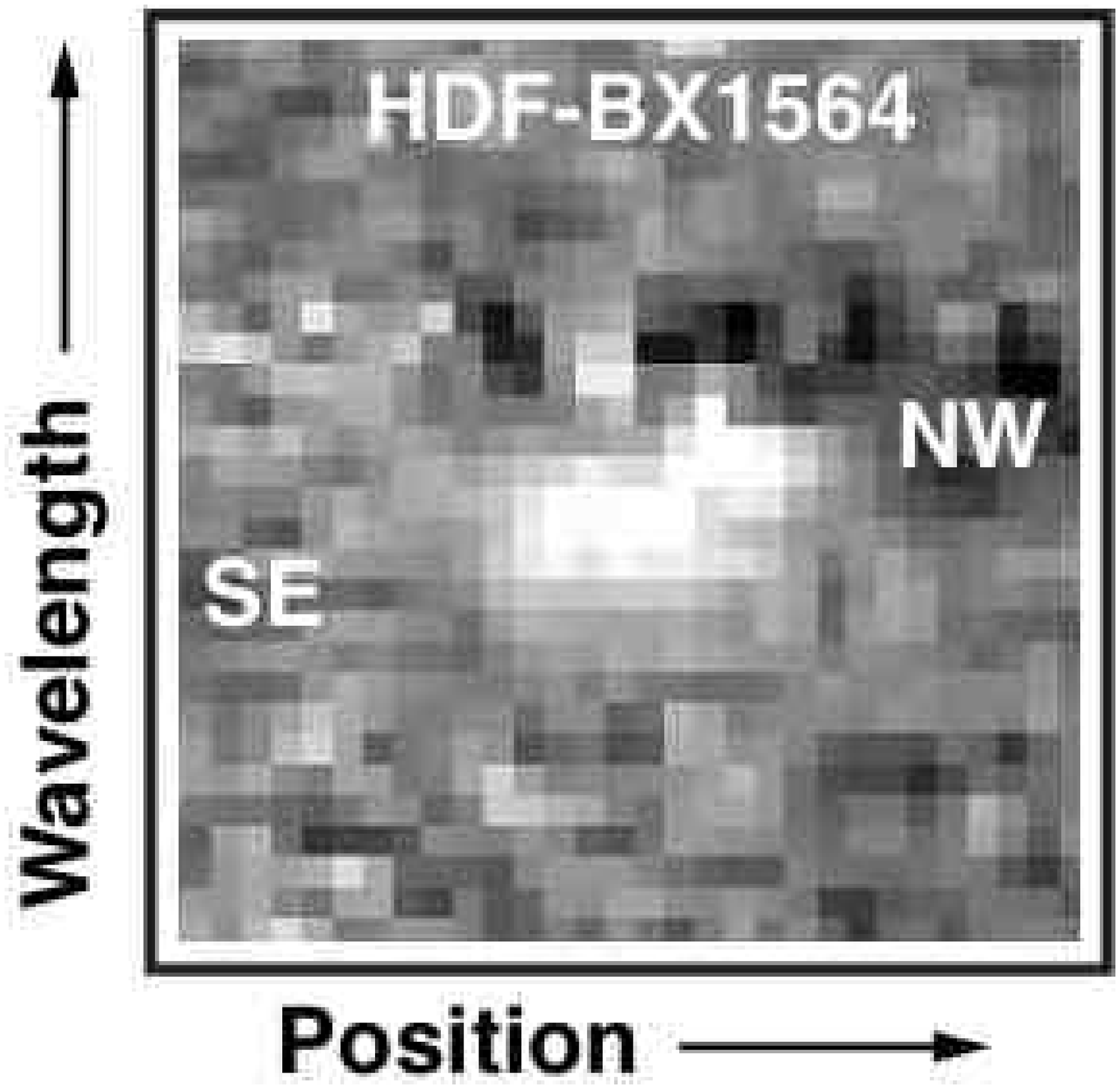}
\caption{Two-dimensional NIRSPEC long-slit spectrum of HDF-BX1564.  Each pixel corresponds
to 0.143'' along the spatial and 4.2$\,$ \AA $\,$ along the spectral axes.  Note the abrupt wavelength transition between SE and NW components.}
\label{bx1564_nspec.fig}
\end{figure}

The local velocity dispersions of the NW and SE features
are $60\pm13$ and $86\pm8$\,\kms\ respectively, consistent with
the typical values
observed for the rest of the galaxy sample, while their effective PSF-corrected radii (\S \ref{gasmorph.sec}) are 1.3 and 0.1 kpc (i.e., the SE feature is consistent with a point
source given the size of the PSF). The respective dynamical masses of the two components are
0.5 and $0.1 \times 10^{10} M_{\odot}$.  
Given the poorly resolved nature of the SE component however, the radius of this object is particularly ill-defined and given the similar values
of $\sigma_{\rm mean}$ for the two
components they may be approximately equal mass.

The multiple components and low surface-brightness features in the UV morphology 
may indicate that HDF-BX1564 is an ($\sim$ equal mass) major merger (although see discussion by Law et al. 2007b)
In this case, additional insight may be gained from 
Keck/NIRSPEC long-slit spectroscopy obtained in June 2004 (see Erb et al. 2006b) with a 0.76'' wide slit aligned along the axis between the two major 
components to within 5$^{\circ}$.
As illustrated in Figure \ref{bx1564_nspec.fig}, 
both primary components are well detected in the spectrum with similar kinematic and spatial separation as in the OSIRIS data.
In this considerably deeper observation there is evidence for nebular emission extending between the two components, 
the SE feature has a spatial FWHM of 1.1'' and reaches all of the way to the NW component, whose
0.6'' FWHM is more closely matched to the PSF of these seeing-limited data.
Notably, there is no gradual shift in wavelength along the axis connecting the two, but instead a sharp transition upon reaching the NW feature, suggesting
that the emitting material between the two components does not partake in a smoothly varying velocity curve and bolstering the case for a major merger
interpretation of this galaxy.

Intriguingly, there is faint detection of \ntwo $\,$ emission ($12 +$ log($O/H$) $= 8.67\pm0.04$) in the OSIRIS data centered on neither of
 the two primary \Ha $\,$ emitting regions, but on a region of space between them
corresponding to a faint spur in the rest-UV morphology (Fig. \ref{hst.fig}).  This may suggest the presence of an obscured AGN at the dynamical center of this system
(although neither the rest-UV spectrum nor the SED shows  evidence for an active nucleus), while the
UV and nebular line emission are dominated by off-center regions of relatively unobscured star formation.
Alternatively, this could be evidence for diffuse shocked gas, or low ionization-parameter gas as discussed by Martin et al. (1997).
Given the low SNR of our data however, it is not possible to conclusively explain the observational complexities of this system.

\subsection{Q1623-BX453}

As discussed in Law et al. (2007a) this galaxy is one of the best examples of the ``typical'' kinematics observed in our sample, with spatially well-resolved
nebular emission exhibiting a negligible velocity differential across the majority of the galaxy and kinematically dominated by a high local velocity dispersion
$\sigma_{\rm mean} = 78 \pm 23$ \kms.  The 
\vsig value calculated for this galaxy is dominated 
by a region of high relative velocity at the northeastern end of the galaxy.  This  region is detected with low confidence however
(SNR $\sim$ 4) and represents an abrupt kinematic discontinuity rather than an extended gradient.

Q1623-BX453 is inferred to have a large amount of cold gas, a sizeable stellar mass ($\sim 3 \times 10^{10} M_{\odot}$),
high metallicity ($12 + $ log($O/H$) $= 8.67\pm0.02$),
and a high SFR surface density with a correspondingly large
outflow velocity $\sim 900$ \kms as traced by rest-UV spectroscopy of interstellar absorption lines.
Given the large quantity of cold gas in this galaxy, it is particularly intriguing that there is no evidence for the rotational structure in which
such gas is traditionally supposed to reside.

\subsection{Q1623-BX502}

Q1623-BX502 is one of the lowest stellar mass objects observed with OSIRIS ($M_{\ast} = 3 \times 10^9 M_{\odot}$) 
and is the only galaxy to be successfully detected under poor observing conditions with high humidity and $V$-band seeing $\gtrsim 1$''.  
These sub-optimal observing conditions are reflected in the 
relatively broad PSF (220 mas) compared to the other galaxies in Figure \ref{gals.fig}.  In addition, Q1623-BX502 was one of two galaxies
which required correction for a spatially distorted PSF as described in \S \ref{datared.sec}.
While there is evidence for resolved velocity structure 
with peak amplitude $\sim$ 50 \kms, this shear is not aligned with any obvious  morphological major axis.
A similar result was derived from the SINFONI observations
of this galaxy by F{\"o}rster Schreiber et al. (2006).

\subsection{Q1623-BX543}

This galaxy consists of two spatially distinct components separated by $v_{\rm 2c} =$ 125 \kms and $d_{\rm 2c} = 6.7$ kpc (0.8'')
in projection (with one almost due north of the other).
Both components are present in our deep ground-based $\cal{R}$-band imaging, which shows a slight
elongation along the N -- S axis, indicating that both components correspond to rest-UV continuum emission regions.
Given the typical $\sim 1''$ PSF of this rest-UV imaging data however it is not possible to determine whether the continuum image consists of a single elongated
feature or two separate features similar to those shown in Figure \ref{gals.fig}.

\begin{figure*}
\plotone{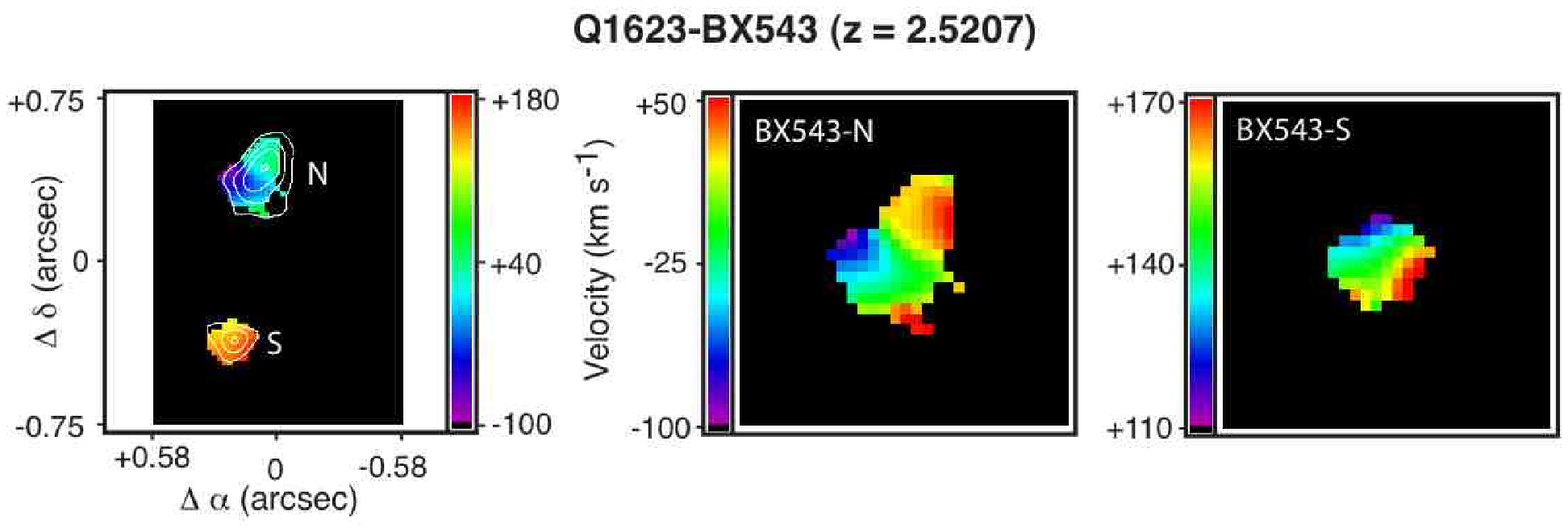}
\caption{The velocity map of Q1623-BX543 (left panel, see also Fig. \ref{gals.fig}) is expanded about each
of the two components labelled North (`N') and South (`S').  The color scale represents line-of-sight velocity relative
to the systemic redshift in all three panels, the limits of this velocity scale vary as indicated to best illustrate
the velocity features of each component.}
\label{bx543_pieces.fig}
\end{figure*}

While not apparent in Figure \ref{gals.fig}
due to the limitations imposed by the dynamic range of the velocity scale, there is velocity structure within the northernmost (N) component.
In Figure \ref{bx543_pieces.fig} we plot an expanded view of each of the components which better shows this structure.  While the general trend is along
a roughly NW-SE axis (see also Fig. \ref{slitcurves.fig}) there are appreciable deviations from a smooth trend along this axis.
We note particularly that the relative velocity of the S component  is strongly inconsistent with a smooth continuation of this velocity field
(i.e. it would need to be {\it blueshifted} relative to systemic rather than redshifted).
Given these discrepancies it is clearly impossible to fit both components of this system with a single simple kinematic model.

Perhaps the most likely scenario is that the primary (N) component is akin to the typical single galaxies observed in the rest of our sample, which in this case
is experiencing a merger  with the smaller S component.
The N/S components have $\sigma_{\rm mean} = 139$ and 60\,\kms,
and effective radii $r = 1.1$ and 0.7\,kpc respectively.
Assuming that each component is relaxed and kinematically dominated by $\sigma_{\rm mean}$ we obtain (Eqn. \ref{dynmass.eqn}) estimates of the
dynamical masses $M_{\rm dyn} \sim 2.5$ and $0.3 \times 10^{10} M_{\odot}$.
If this system is a merger, these rough
calculations suggest that the mass ratio of the merger is $\sim 8/1$.

While it is somewhat curious that the photometry of this source is (relatively) well-fit by a simple stellar population model, such a mass ratio suggests that the larger of the two
components may constitute the bulk of both the current and past star formation in the system.  The total stellar mass of the system is quite
small ($4 \times 10^9 M_{\odot}$) with an extremely young stellar population as might be expected if rapid star formation\footnote{The high 
SED-derived SFR given in Table \ref{SED.table} for Q1623-BX543
and Q1700-BX490 are likely significantly overestimated, and the 10 Myr ages highly uncertain due to difficulties fitting appropriate $E(B-V)$ for
these very young galaxies (see discussion in Reddy et al. 2006b).}
is occurring as a result of the rapid injection of large quantities of cold gas.

\subsection{Q1700-BX490}

As in the case of Q1623-BX543, this galaxy has a complex, multiple-component structure.
As demonstrated by Figure \ref{hst.fig}, the overall location and size of these two components is broadly consistent with \hst $\,$ rest-UV imaging.  At the fainter
surface brightness levels probed by \hst $\,$ both components are enclosed in a common envelope of flux, and it is probable that the discontinuous \Ha\ morphology
is simply a consequence of our limiting surface brightness sensitivity.  Curiously, the W component has a single well-defined
peak to the UV flux distribution located between the double peaks of nebular \Ha $\,$ emission (see contours in Fig. \ref{hst.fig}).

In the larger western (W) component there is a smoothly varying velocity gradient along the NE -- SW axis (i.e. not aligned with the E-W major axis
of the \hst $\,$ morphology) from which emission line flux appears
to be concentrated in two closely separated knots.  Rather than peaking in the center of the object as might be expected for a rotating gaseous disk
(although see discussion by Wright et al. 2009),
the velocity dispersion instead peaks in one of these two knots.
While the smaller eastern (E) component is close to the main body in projection (3.4 kpc) and would likely appear to  connect with the W component in deeper spectroscopic data,
it does not partake in the velocity gradient defined by the W body.  Rather, it has its own (albeit weak) velocity gradient aligned nearly perpendicular ($\sim 70^{\circ}$) to that of
the main body.

Paired with the additional faint structure evident in the \hst $\,$ image (note particularly another low surface-brightness component SW of the main body) one natural explanation for this galaxy
may be that it represents a merger (mass ratio $\sim$ 5/1)  with multiple sites of star formation.
The broadband photometry is extremely well fit by a young (10 Myr) stellar population model with a small stellar mass ($4 \times 10^9 M_{\odot}$) and
a high current star formation rate.
This galaxy is also likely quite metal-poor
as \ntwo $\,$ is undetected, placing a limit
$12+\log({\rm O/H}) \leq 8.26$ ($\leq 2/5$ solar).

\subsection{Q1700-BX710}

As for Q1623-BX453, this galaxy is relatively well-resolved with no strong evidence for spatially resolved velocity structure.  
While \ntwo $\,$ is detected in emission, it is weak (12 + log($O/H$) $= 8.40\pm0.08$) and consistent with a point source roughly concentric with the \Ha $\,$ flux peak.
There is a slightly redshifted extension of flux to the northwest of the galaxy core
which has a lower velocity dispersion and is coincident with the low surface brightness
emission tail seen in \hst $\,$ imaging data (Fig. \ref{hst.fig}).
We note that another galaxy (Q1700-BX711) is located $\sim$ 5'' (41 kpc) to the northwest of Q1700-BX710 (i.e., approximately along the direction of
the `tail' indicated by the \hst $\,$ data) and at the same redshift to within 10 \kms.
However, a long-slit NIRSPEC spectrum obtained along the line connecting the two galaxies shows no evidence for star formation occuring 
in the region between the two galaxies.  If these galaxies are genuinely close in space as implied by their redshifts they may merge over the next few hundred Myr.

\subsection{Q1700-BX763}

Q1700-BX763 has an apparent western spur to the \Ha $\,$ flux consistent with that observed in \hst $\,$ imaging data.
While there is mild evidence for a velocity gradient along this spur the magnitude of this gradient is small and its significance low (Fig. \ref{slitcurves.fig}).
As for Q1700-BX710, weak \ntwo $\,$ emission is detected (implying 12 + log($O/H$) $= 8.41\pm0.25$)
that is consistent with a point source and is roughly (within $\sim 0.1''$) concentric with the \Ha $\,$ emission peak.

\subsection{DSF2237a-C2}

DSF2237a-C2 is located at a significantly higher redshift than any other galaxies in the target sample ($z \sim 3.3$), and is drawn from the ``LBG''
galaxy catalog of Steidel et al. (2003).
Given its high redshift, it is interesting that DSF2237a-C2 has some of the strongest velocity shear observed in our sample ( $\vsig =0.6 \pm 0.2$)
and is the only isolated case where this shear is consistent with rotation and unambiguously aligned with the morphological major axis.
Nonetheless,  the local velocity dispersion of DSF2237a-C2 is comparable to that of the rest of the galaxy sample with
$\sigma_{\rm mean} = 89\pm20$ \kms. 
We note that this galaxy was discussed previously by Law et al. (2007a), although various estimate of the kinematics and flux parameters have
been updated here for consistency with our new calibration routines.\footnote{In particular, the value of $L_{\othree}$ in Table 3 of Law et al. (2007a) was 
incorrectly given as $2.2 \times 10^{42}$ erg s$^{-1}$, while it
should instead have been given as $22 \times 10^{42}$ erg s$^{-1}$.  The revised number is consistent with the estimate 
of $16.8 \times 10^{42}$ erg s$^{-1}$ derived using our new flux calibration routine.}


\subsection{Q2343-BX418}

As for Q1217-BX95, Q2343-BX418 is well detected in the composite OSIRIS data cube but is spatially compact
with a slight elongation along the N--S axis.  While there is some evidence for coherent velocity structure aligned with the morphological minor axis,
the magnitude of this shear is significantly smaller than the local velocity dispersion.
We note additionally that this is our lowest mass galaxy with $M_{\ast} = 1 \times 10^9 M_{\odot}$,
a similarly small dynamical mass estimate $M_{\rm dyn} = 3 \times 10^9 M_{\odot}$ and very little dust ($E(B-V) = 0.03$).
With so little extinction, Q2343-BX418 is also a strong Ly$\alpha$ emitting source (see Pettini et al. {\it in prep.}).

\subsection{Q2343-BX513}

\begin{figure}
\plotone{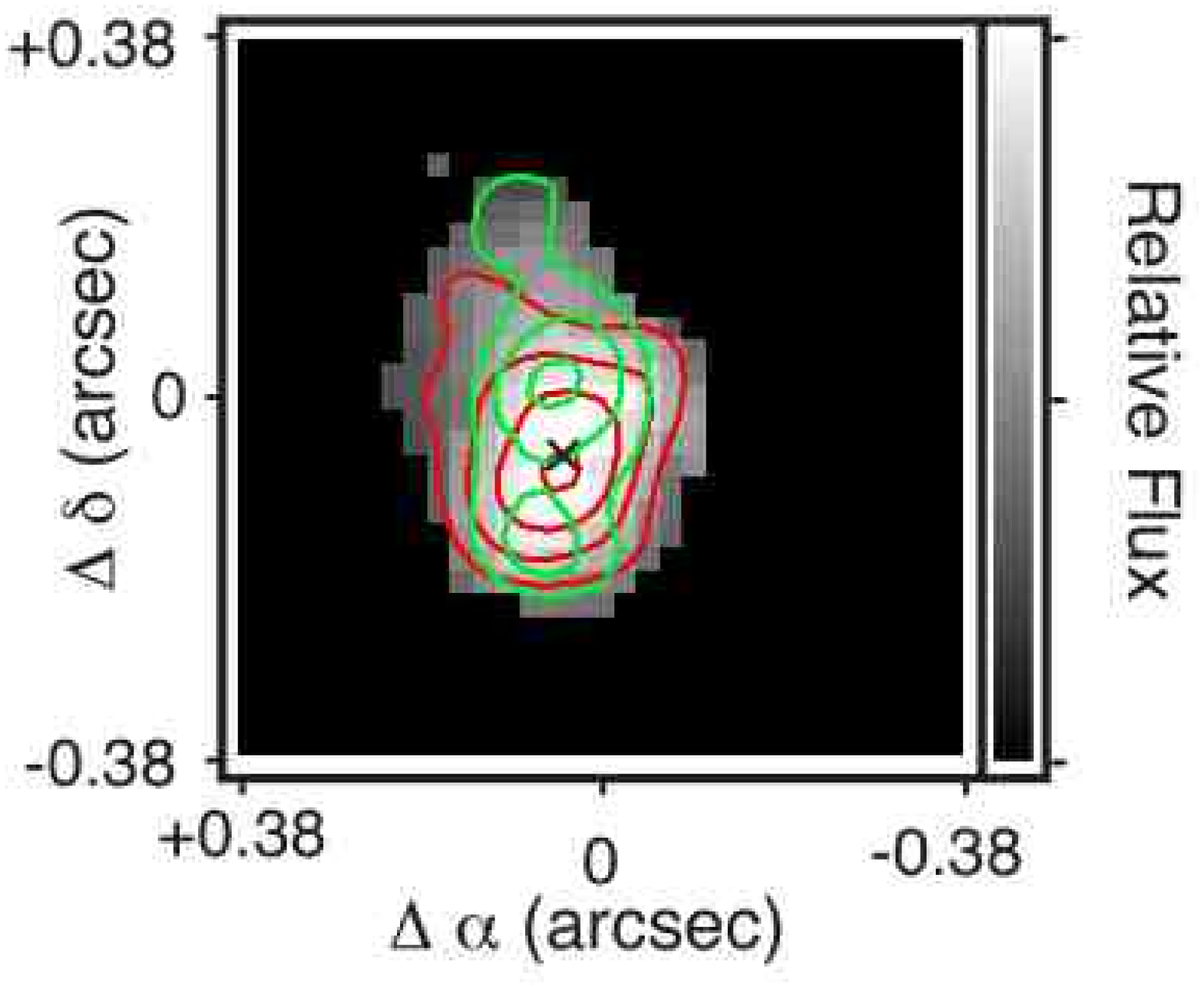}
\caption{Flux map of Q2343-BX513 overlaid with contours showing the distribution of flux in the blue and red sides of \Ha $\,$ emission (green and red contours respectively).
The wavelength difference between the two sets of contours is 10 \AA, or 150 \kms at the redshift of the galaxy.  The black cross indicates the location of peak \ntwo $\,$ emission.}
\label{bx513cont.fig}
\end{figure}

Q2343-BX513 is elongated along the N--S axis and exhibits a complex  velocity structure composed of multiple spatially superimposed components.  
At first glance the one-dimensional velocity curve in Figure \ref{slitcurves.fig}
(with $\vsig = 0.7\pm0.2$) gives an indication of smooth resolved
velocity shear in this source, but
the two-dimensional velocity map (see Figure \ref{gals.fig})
is inconsistent with a simple rotational model.
Following the velocity pattern from North to South, we find
that the northernmost region is blueshifted by $-80$\,\kms
(relative to the systemic redshift); moving South
we reach a maximum redshift of $v = +50$\,\kms
near the flux peak, but then the velocity falls off again
to lower values, down to $v \simeq 0$\,\kms
at the southern edge of the galaxy.
Indeed, if we consider the spatial distribution of the red and blue sides of the \Ha $\,$ emission individually we note that while the redder component is well-centered
the blue edge of \Ha $\,$ emission is dominated by two distinct emission regions
located directly north and south of the center (Fig. \ref{bx513cont.fig}).
The total \Ha $\,$ flux of Q2343-BX513 is dominated by the centrally concentrated redshifted component.

This double-component morphology helps to explain the previous long-slit observations of Erb et al. (2006c) who found two slightly different redshifts for the system
with different $\sigma$ when the galaxy was observed at two different position angles.
Notably, the galaxy also has a highly unusual double-featured UV spectrum with Ly$\alpha$ emission at redshifts $z = 2.106$ and 2.114 (i.e. offset by 770 \kms)
and possible doubles of the strongest
interstellar absorption lines, such as
\sitwo\,$\lambda 1260$, \oone\,$\lambda 1302$,
and \ctwo\,$\lambda 1334$.
This spectrum is additionally confused by the presence of absorption lines arising in an intervening system Q2343-MD80 ($z = 2.014$)
located a few arcseconds away.
\ntwo\,$\lambda 6583$ is detected with a strength relative to
\Ha\ which implies $12+\log({\rm O/H}) = 8.63\pm0.03$,
close to the solar oxygen abundance,
and is roughly concentric with the redder \Ha $\,$ component with a similar north-south elongation.
There is a significant velocity gradient in the \ntwo $\,$ emission  with an end-to-end velocity differential $\sim$ 110 \kms.

We note that this galaxy is the oldest in the OSIRIS sample (among both detected and undetected targets) with an estimated stellar population age of $\sim $ 3 Gyr and a large
stellar mass $M_{\ast} = 7.8 \times 10^{10} M_{\odot}$.  One possibility, therefore may be that the redshifted component (with associated \ntwo $\,$ emission and the larger
velocity dispersion) is the massive ``mature''
galaxy which contains the majority of the evolved stellar population and has an appreciable velocity gradient (as expected for such a massive galaxy; see discussion in 
\S \ref{disc.sec}).  Behind this component may be a smaller (infalling) system which is partially obscured by the central core of the foreground galaxy, causing the apparent
two-component morphology.
While it is beyond the scope of the present work to verify this hypothesis (and indeed, the two components may simply be unrelated juxtapositions in redshift space),
we may certainly conclude that this system is particularly complex.

%

\subsection{Q2343-BX660}
\label{q2343bx660.sec}

The  \Ha $\,$ morphology of Q2343-BX660 is composed of  two spatially distinct peaks (which are more apparent in the 3-d data cube than in the collapsed \Ha\ map)
which blend together at low surface brightness to form a contiguous system.
While there is a strong velocity differential across the combined system, 
the line-of-sight velocity remains relatively constant across each of these two clumps and changes rapidly in the region between the two
($\sim 80$ \kms over less than 0.1''). 
While turnovers in the velocity profile are naturally found in rotating systems beyond a certain
radius, it is suspicious that the entire velocity gradient of this galaxy is contained within a single angular resolution element (corresponding to a radius  $\sim$ 400 pc).

Instead, this figure is reminiscent of Figure 8 in Law et al. (2006), whose panel (d) shows the nearly identical velocity map expected for a 2-clump velocity model in which
the apparent scale of the `turnover' is determined by the observational PSF blurring together emission from the two regions.
Indeed, Nesvadba et al. (2008; see also Wright et al. 2009) recently presented observations of a galaxy at $z \sim 3$ (Q0347-383 C5) which showed a very similar velocity structure, transitioning
rapidly from a plateau of $\sim 60$ \kms to $\sim -80$ \kms within a negligible physical distance, leading the authors to favor a merger interpretation for the galaxy.
Similarly, we conclude that the kinematics of Q2343-BX660 are more consistent with merging compact galaxies than with underlying rotation in a single relaxed system.
Given the relative constancy of $\sigma$ across the galaxy we surmise that in such an event the mass ratio should be $\lesssim 3/1$, with a total stellar mass of 
$\sim 10^{10} M_{\odot}$.

\section{DISCUSSION}
\label{disc.sec}

\subsection{Selection Effects}
\label{selfx.sec}

As outlined in \S \ref{tarsel.sec}, targets were selected from the rest-UV galaxy sample
for a wide variety of reasons (high/low stellar mass, extended/compact morphology, presence/absence
of kinematic shear in long-slit spectra, etc.) subject
to the general criterion that we believed the galaxy would be detected on the basis 
of previous long-slit observations (Erb et al. 2006b) or narrowband \Ha\ imaging.
While our original choice of targets attempted to
include a wide range of galaxy properties, it is still
difficult to know how representative our results are
of the general population of star-forming galaxies at
$z = 2 - 3$, given the large fraction of targets
which were undetected or poorly detected (11 out of 24) despite extensive observational efforts (see Table \ref{targets.table}).
This bias is difficult
to quantify since the probability of detecting a given galaxy is a strong function of observing conditions; 11 of the 16 unsuccessful 
observations (some of which represent
multiple attempts to observe the same galaxy) listed in
Table \ref{targets.table} were performed under sub-optimal conditions typically characterized by poor seeing $\gtrsim 1''$ (resulting in poorer AO correction), 
high humidity (exacerbating telluric absorption), and/or attenuation by cirrus ($\gtrsim 1$ mag).
In contrast, only one of our successfully detected
galaxies (Q1623-BX502)  was observed under similarly poor conditions.

\begin{figure}
\plotone{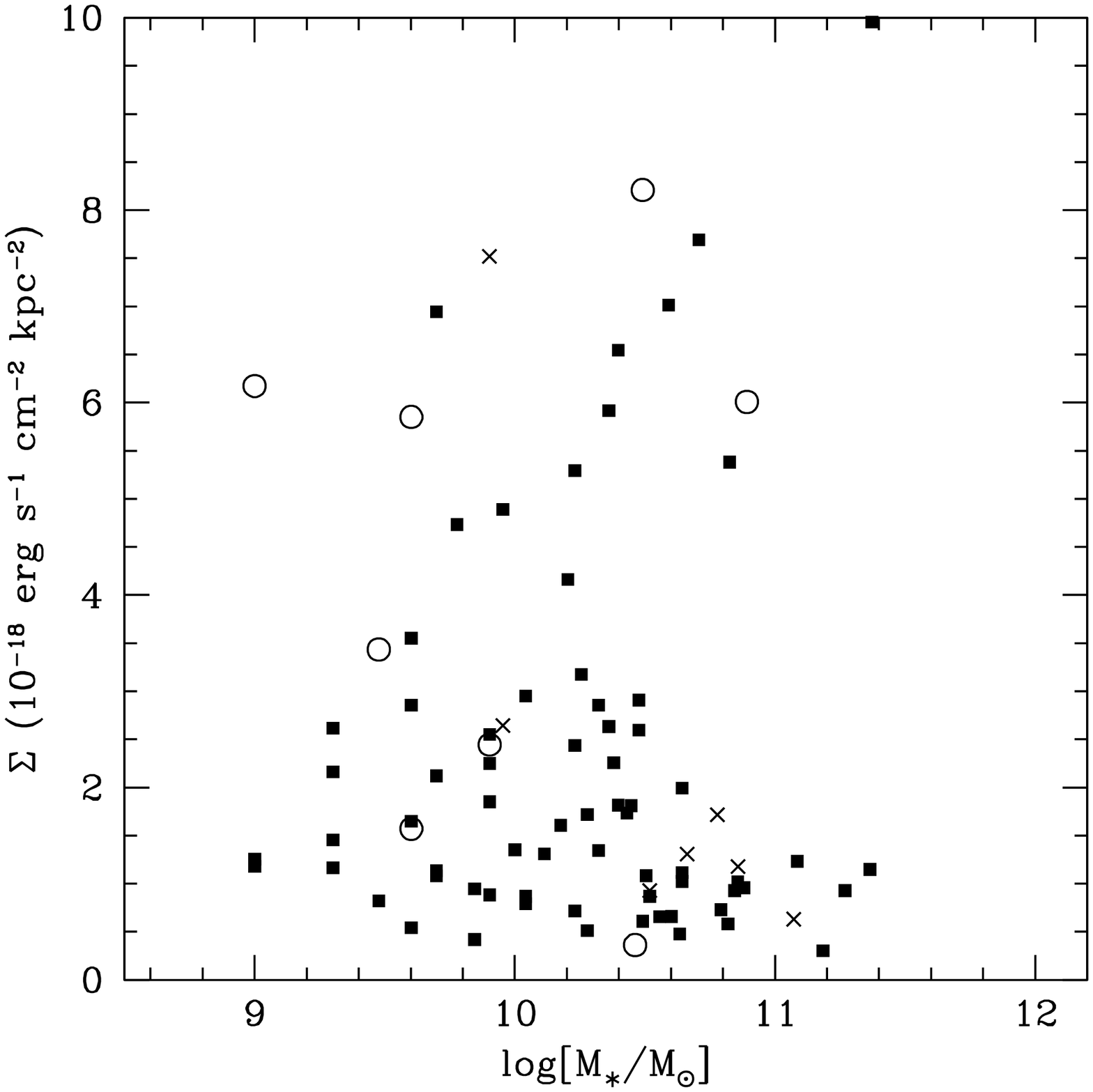}
\caption{Mean \Ha\ surface brightness as a function of stellar mass for the Erb et al. (2006c) long-slit survey of galaxy kinematics.  Data are plotted for all galaxies in the Erb et al. (2006c) survey in the redshift range $z = 1.9 - 2.5$ which have well-defined values of both the \Ha\ flux and galaxy size.  Open circles represent galaxies which have been successfully observed
with OSIRIS, crosses galaxies which were either undetected or poorly detected with OSIRIS, and filled squares galaxies not observed with OSIRIS.}
\label{mstar_surfb.fig}
\end{figure}

The likelihood of detection was probably also governed by the \Ha\ surface brightness of the target galaxies.
In Figure \ref{mstar_surfb.fig} we plot the surface brightness $\Sigma = F_{\Ha}/ r_{\Ha}^2$
of galaxies targeted by the Erb et al. (2006c; based on values given in their Table 4) 
long-slit spectroscopic survey, indicating galaxies which were detected/undetected in the OSIRIS observations.
While the results naturally will be quite inhomogenous given the widely varying weather conditions and exposure times of the OSIRIS observations we note that
the mean surface brightness of galaxies in the NIRSPEC survey which were successfully detected with OSIRIS 
is $4.3 \times 10^{-18}$ erg s$^{-1}$ cm$^{-2}$ kpc$^{-2}$,
compared to $2.3  \times 10^{-18}$ erg s$^{-1}$ cm$^{-2}$ kpc$^{-2}$ for galaxies either undetected or poorly detected with OSIRIS, and a mean/median of 
2.6/1.7 $ \times 10^{-18}$ erg s$^{-1}$ cm$^{-2}$ kpc$^{-2}$ respectively for the entire NIRSPEC sample.
Clearly, we are observing a population of galaxies with somewhat higher than average star formation rate 
surface density.
 
Such a surface-brightness limit will also bias our observations towards high surface-brightness regions within a given galaxy.
Indeed, evidence suggests that at least some of the observations may genuinely be missing some flux from fainter surface-brightness regions.  
First, the integrated \Ha\ fluxes are about half 
the aperture-corrected values determined from long-slit spectroscopy by Erb et al. (2006b), although in some cases this may also indicate that the aperture correction adopted
for the long-slit data was overestimated.
Similarly, galaxy sizes estimated from the long-slit program tend to be somewhat larger (although considerably less certain given the seeing-limited PSF) than those from OSIRIS.  
Additionally, the dynamical mass estimates are considerably less than the total
of the stellar + gas masses, as might be expected if the radius of the system is underestimated in Eqn. \ref{dynmass.eqn} (or the stellar and/or gas masses are distributed over a
larger volume than the ionized gas traced by OSIRIS).

In some cases (e.g. DSF2237a-C2) the kinematics suggest that the rotation curve may continue to rise at larger radii, meaning that deeper spectroscopy could result in larger
values for $v_{\rm shear}$.  This is unlikely to be the case in general however, since at least at least six of the 13 galaxies are consistent with a velocity gradient
of $\frac{dv}{dx} = 0 $ km s$^{-1}$ kpc$^{-1}$.  Indeed, deeper imaging of one galaxy (Q1623-BX502) with VLT/SINFONI (F{\"o}rster Schreiber et al. 2006) 
showed no evidence for significant flux or velocity shear beyond that seen with OSIRIS.

Figure \ref{mstar_surfb.fig} also suggests that we are more likely to detect younger galaxies than older galaxies with a larger established stellar mass.
The 13 galaxies detected with OSIRIS have a mean
assembled stellar mass
$\langle \log (M_{\ast}/M_{\odot}) \rangle  = 10.1$,
compared to the mean
$\langle \log (M_{\ast}/M_{\odot}) \rangle  = 10.6$ (i.e., three times more massive)
of the eleven galaxies which were undetected (or only poorly detected) in the OSIRIS data.
Thus, while our sample may be representative of typical-mass galaxies in the star-forming population
(the full sample of 818 optically-selected galaxies at redshifts $1.8 < z < 2.6$ with stellar mass information in our various survey fields has a mean stellar mass
$\langle \log (M_{\ast}/M_{\odot}) \rangle  = 10.07$; Fig. \ref{mstartrend.fig}; see also Reddy et al. 2006a),
the observations presented here are not particularly 
sensitive to the high-mass end of the star-forming galaxy distribution.
We note for completeness however that all of these masses are high relative to the {\it total} galaxy population
(i.e., including both UV-bright and faint galaxies).
Integrating the stellar mass function given by Reddy et al. (2009) suggests that while galaxies in the stellar mass range probed by OSIRIS
(i.e., $0.1 - 7.8 \times 10^{10} M_{\odot}$) constitute 
$\sim 66$\% of the total space density of galaxies at $z \sim 2$, 90\% of the total population have masses smaller than the average of the OSIRIS sample.

\subsection{Expanding the Sample}
\label{otherobs.sec}

In order to construct a more complete picture
of the overall $z \sim 2$ galaxy population
we therefore combine our results with those of
similar studies recently undertaken by a variety of authors.
The largest sample of IFS observations of galaxies
at redshift $z \sim 2$ to date is the
``SINS'' survey using SINFONI on the VLT,
the results of which have been presented
by F{\"o}rster Schreiber et al. (2006);
Genzel et al. (2006);  Bouch{\'e} et al. (2007);
and more recently Shapiro et al. (2008) and Genzel et al. (2008).\footnote{We refer the reader also to F{\"o}rster Schreiber et al. (2009),
which first became available during the publication process of this paper.}
While all of the SINS galaxies (a few of which are
in common with the present survey) exhibit comparably high
velocity dispersions to the 13 OSIRIS galaxies reported here,
many of them also show ordered velocity shear $\gtrsim 100$ \kms to which
Genzel and collaborators (see particularly Shapiro et al. 2008) have been relatively
successful in fitting turbulent disk models.

In part, the different prevalence of shear between the SINS and OSIRIS surveys is a consequence of their sensitivity.
Due to a combination of optical design differences and a larger (0.1'') spaxel scale
SINFONI is roughly twice as sensitive as OSIRIS and therefore able to probe fainter surface brightness
features to larger radii.  
Combined with longer integration times
(18000\,s. vs. 8100\,s in the case of Q2343-BX389, for example),
SINFONI is capable of
detecting large-scale velocity gradients
in lower surface-brightness galaxies undetectable with OSIRIS.
In contrast, the image quality of the OSIRIS data is generally superior, with LGSAO
image correction for all of the target galaxies and a typical PSF $\sim$ 100 mas, compared to the large fraction of the SINS sample (13 of the 14 galaxies discussed by
 F{\"o}rster Schreiber et al. 2006)
which were obtained in seeing-limited mode with a PSF $\sim$ 500 mas (with some notable exceptions; e.g., Genzel et al. 2006).
One simple explanation may therefore be that velocity shear is less noticeable in the inner few kpc of $z \sim 2$ star-forming galaxies
(where the OSIRIS survey is most sensitive) and only becomes significant relative to the line-of-sight velocity 
dispersion at larger radii and fainter surface brightnesses
(where the SINS survey is most sensitive).  Indeed, the maximum kinematic {\it slope}  of DSF227a-C2 (the galaxy which presents the strongest case
for resolved velocity shear in an inclined system for the OSIRIS sample) $\frac{dv}{dx} \sim 35$ km s$^{-1}$ kpc$^{-1}$ is comparable to that of the weakest
rotator in the Genzel et al. (2008) sample (SSA22a-MD41), and if this slope extends to larger radii the two galaxies may also have comparable $v_{\rm shear}$ when observed 
to a similar limiting surface brightness.

In general however, such a purely instrumental bias between the two surveys is an unsatisfactory explanation for the observed kinematic differences.  While the upper end
of the kinematic slopes observed with OSIRIS overlaps with that of the SINFONI ``massive disk'' population, the mean value 
for the 13 galaxies presented here ($\langle \frac{dv}{dx} \rangle = 19$ km s$^{-1}$ kpc$^{-1}$) 
is substantially less than the mean ($\langle \frac{dv}{dx} \rangle = 47$ km s$^{-1}$ kpc$^{-1}$) of the 5 
galaxies discussed by Genzel et al. (2008).  Indeed, 6 of the 13 galaxies observed with OSIRIS have $\frac{dv}{dx}$ consistent with 0 to within observational uncertainty.
In addition, both SINFONI and OSIRIS find similar kinematics
in the few galaxies (e.g. Q1623-BX502) which were successfully observed by both surveys and do not indicate substantial shear in a low surface-brightness component.

Most likely, the  kinematics probed by the two surveys represent a continuum ranging from genuinely dispersion-dominated to rotationally supported systems.
This range in kinematic properties is perhaps unsurprising in light of the relative physical properties of the galaxies included in the two samples.
As discussed above, OSIRIS targets were drawn from the opticaly-selected ``BX'' 
galaxy sample; the successful observations tend to have stellar masses in the less-massive to typical-mass range. 
In contrast, many of the SINS galaxies with significant velocity shear have inferred stellar masses that would place them in the top quartile of the general ``BX'' sample. 
F{\"o}rster Schreiber  et al (2006) discuss a sample of 12 UV-selected galaxies (excluding Q1307-BM1163 due to its lower redshift, and Q1623-BX663 due to 
its subsequent identification as an AGN) with a similar range in $M_{\ast}$ compared to the OSIRIS sample. The four galaxies in the sample with unambiguous 
velocity gradients with $v_{\rm shear} \gtrsim 100$ \kms 
(SSA22-MD41, Q2343-BX389, Q2343-BX610, Q2346-BX482)
have an average stellar mass $\langle $log$(M_{\ast}/M_{\odot})\rangle  = 10.99$, 8 times greater than the average of the rest-UV selected 
population.\footnote{Where available, we have
used stellar masses for the SINS galaxies calculated by Genzel et al. 2008.  Otherwise, masses are drawn from our own stellar population modeling.  Both
estimates are consistent and adopt a Chabrier (2003) IMF.  For our galaxies with more than one component we focus on the velocity structure of the most massive component.}
In their detailed discussion of disk kinemetry, Shapiro et al. (2008) compare to a similarly high mass subset of 11 galaxies chosen for their particularly high quality data.
The mean stellar mass of their 5 rest-UV selected galaxies is $\langle$log$(M_{\ast}/M_{\odot})\rangle = 10.9$, the remaining 6 are selected according to their rest-frame optical
$BzK$ color (e.g. Daddi et al. 2004).  
While there is a large amount of overlap between the BX and $BzK$ samples, the 
particular $BzK$ objects selected all have bright $K$ band magnitudes ($K_s < 20.5$) whereas the BX objects (selected without regard to $K$ magnitude)
extend considerably fainter, to $K_s \sim 22$ (see discussion by Reddy et al. 2006b).
The $BzK$ sample also tend to be slightly less strongly nucleated with fewer spatial irregularities (see discussion by Law et al. 2007b), consistent with the interpretation that they might more
accurately be described as evolved stellar ``disks'' than the BX sample.
The Bouch{\'e} et al. (2007) sample in turn is roughly a superset of the F{\"o}rster Schreiber et al. (2006)
rest-UV selected galaxies, $K$-bright rest-optically selected galaxies, and submillimeter galaxies (e.g. Chapman et al. 2005).
In turn, Genzel et al. (2008) focus on five primary galaxies (four previously discussed by F{\"o}rster Schreiber et al. (2006) and one additional galaxy BzK6004) with a mean
stellar mass $\langle$log$(M_{\ast}/M_{\odot})\rangle = 11.0$, 8.5 times larger than the mean of the rest-UV selected population, and more massive than $> 99$\% of the
overall $z\sim 2$ galaxy population (based on the mass function of Reddy et al. 2009).

\begin{figure*}
\plotone{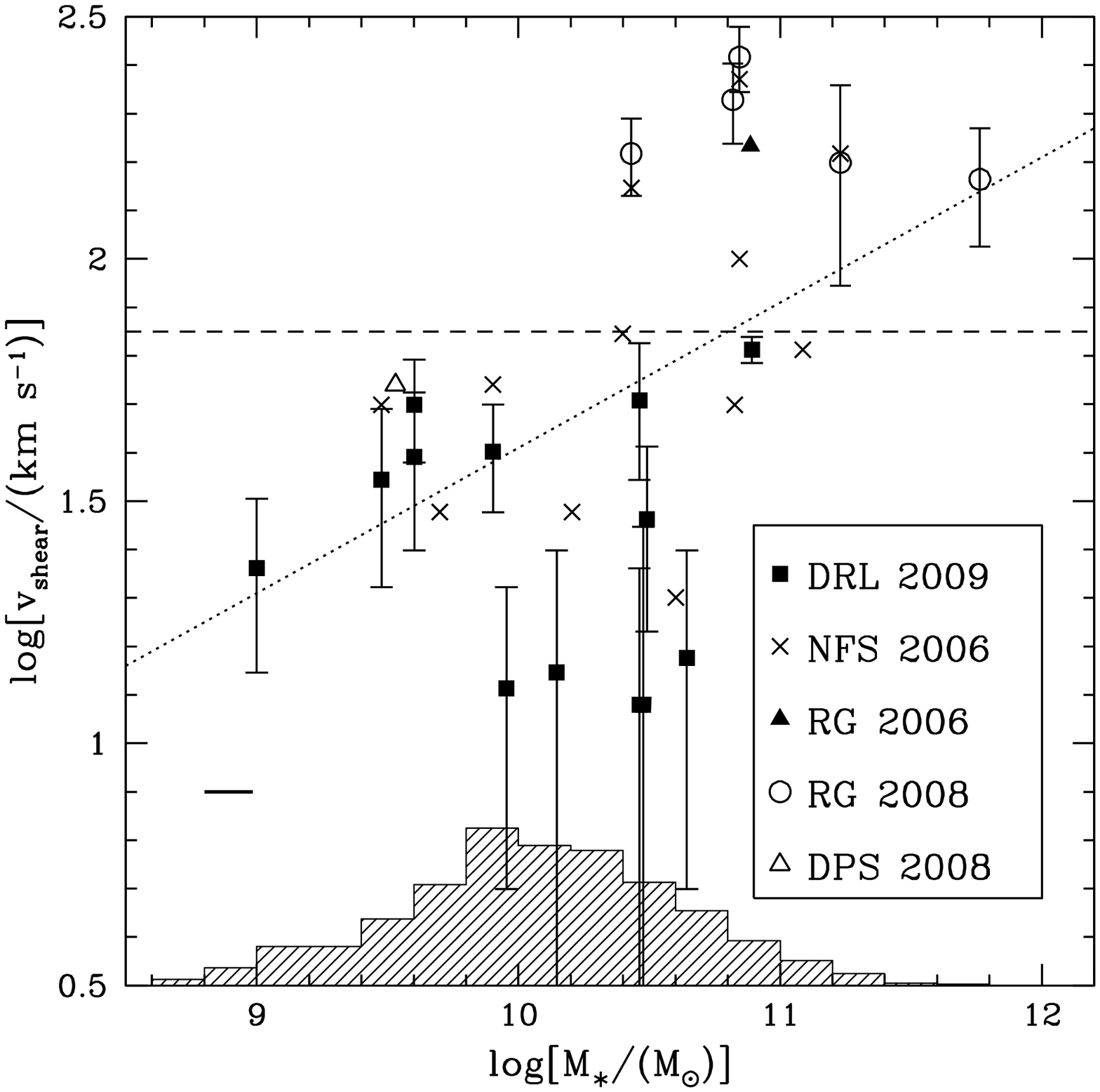}
\caption{Logarithmic plot of stellar mass vs. maximum line-of-sight velocity shear for our 15 galaxies (DRL 2009) and galaxies discussed by 
F{\"o}rster Schreiber et al. (2006; NFS 2006), Genzel et al. (2006; RG 2006), Genzel et al. (2008; RG 2008), and Stark et al. (2008; DPS 2008).  
All data shown here are {\it uncorrected} for inclination effects.
The dashed line denotes $\langle$log$(\sigma_{\rm mean})\rangle$ of our sample (note that most galaxies fall below this line), the dotted line represents a simple least-squares fit to the
data (log ($v_{\rm shear}$/\kms) $= 0.30 \times$ log($M_{\ast}/M_{\odot}$) - 1.39).  The histogram in the lower part of the plot (arbitrary scale) represents the relative number of galaxies
in each logarithmic mass bin for the full sample of 818 optically-selected galaxies with stellar mass information in the redshift range $1.8 < z < 2.6$.  
Uncertainties in $v_{\rm shear}$ are given where available,
the solid line in the lower left corner of the plot indicates the typical uncertainty in $M_{\ast}$.}
\label{mstartrend.fig}
\end{figure*}

These data are summarized in Table \ref{otherwork.table} and Figure \ref{mstartrend.fig}, which plots the maximum observed shear velocity of galaxies as a function of stellar mass.
While there is considerable scatter in the diagram (unsurprisingly given that $v_{\rm shear}$ is not, and in most cases cannot, be corrected for inclination) 
we note that {\it all} galaxies with $v_{\rm shear}$ greater than the typical velocity dispersion have masses $2 - 20$
times greater than the average for rest-UV selected galaxies at these redshifts.

If velocity shear is indicative of systematic rotation, this suggests that stable rotation may be more prevalent (or at least more easily observed)
among galaxies which have already accumulated a sizeable stellar population.
It may be that the lowest mass galaxies tend to display negligible velocity shear and extremely high \Ha $\,$ surface brightness,
while star formation in more massive galaxies with appreciable old stellar populations tends
to trace the kinematics of any underlying disks with greater fidelity.  
While this trend might hold in a {\it statistical} sense, however, we stress that individual galaxies do not always fit this relation:
one of our highest mass objects (Q2343-BX453) is perhaps
the single best example of a galaxy without any resolved large-scale kinematic structure, while some low mass objects (Q1700-BX490) show obvious structure.
Indeed, the observation by Stark et al. (2008) of a gravitationally lensed galaxy offered an exceptionally fine-scale (resolution $\sim$ 100 pc) view of a low-mass galaxy
whose stellar mass and velocity structure are remarkably similar to that of Q1700-BX490 (Table \ref{otherwork.table}) with smoothly varying velocity shear.

\subsection{Local Analogues and Galaxy-Galaxy Mergers}
\label{local.sec}

Recent studies of the clustering properties of $z \sim 2$ star-forming galaxies in conjunction with large cosmological N-body simulations (e.g. Springel et al. 2005)
have indicated (Conroy et al. 2008; Genel et al. 2008) that these galaxies evolve into a variety of galaxy types by the present day, including  typical $L^{\ast}$ galaxies.
Local galaxies with similar star formation properties, however, tend to be gas-rich mergers with multiple nuclei and tidal features
(e.g., Sanders et al. 1988; Bushouse et al. 2002), such as the ULIRGS Mrk 273 and IRAS 15250+3609.
Similar to the high-redshift sample, these galaxies show large velocity dispersions with little resolved kinematic substructure
in the regions of brightest \Ha $\,$ emission, despite relatively strong ($\sim 200$ \kms) and highly disturbed features in lower surface brightness regions
(Colina et al. 2005).
Perhaps the best analog of these high-redshift galaxies that has been found to date however is 
the supercompact subsample of UV-luminous galaxies (ScUVLGs) discovered using
GALEX (Heckman et al. 2005).
These systems have a similar range of specific SFRs, metallicities, and dust content to high-redshift
Lyman Break Galaxies (see discussion by Heckman et al. 2005; Hoopes et al. 2007; Basu-Zych et al. 2007), and recent
evidence (Basu-Zych et al. 2009) indicates that their kinematics are also similar to those of $z \sim 2 - 3$ galaxies
with high $\sigma_{\rm mean} \gtrsim 80$ \kms and $\vsig \lesssim 1$.
Many of these ScUVLGs show tidal features and other indications of active mergers on small physical scales 
and at low surface brightnesses which would not be detectable
at high redshift, leading Overzier et al. (2008) to conclude that rapidly star forming galaxies such as our rest-UV selected sample are mergers of gas-rich galaxies
which have triggered super starbursts on scales $\sim$ 100 - 300 pc.

Numerous morphological studies (e.g., Conselice et al. 2003; and references therein) have interpreted the irregular morphologies of $z \sim 2$ galaxies as evidence
for galaxy-galaxy mergers in the high-redshift universe, and our own kinematic data may further bolster this interpretation in some cases (e.g., HDF-BX1564, Q1700-BX490).
One natural suggestion therefore is that the observed multiple-clump morphologies and unusual kinematics of these galaxies may be a consequence of merger activity
which prevents the formation of a stable disk at early times (e.g., Ostriker 1990).
Such lumps can disrupt  the preferred kinematic axis of disk galaxies in the local universe (e.g., Benson et al. 2004), and their effects would likely be stronger
at higher redshifts when dark matter haloes were merging more rapidly (e.g., Zentner \& Bullock 2003; Fakhouri \& Ma 2008).
By inflating the velocity dispersion of putative disks in all directions (and not simply perpendicular to the disk), such mergers with either luminous or dark satellites may help mask
weak rotational signatures while more massive galaxies with correspondingly larger rotational velocities may naturally be more resilient to such disruption.

However, comparison with cosmological merger trees suggests (Conroy et al. 2008; Genel et al. 2008) that major mergers are too rare to account for the high 
space density of all $z \sim 2$ star-forming galaxies.  In addition, some of our best
merger candidates are simultaneously those with the clearest rotation curves (Q1700-BX490).
It is therefore unlikely that the uniformly high velocity dispersion of {\it every single star-forming galaxy}
observed to date at redshift $z \sim 2$ is due to large-scale merger activity, suggesting that the star-formation may instead be due to a different triggering mechanism
than observed in the local universe.

\subsection{An Evolving Picture of Galaxy Formation}
\label{theory.sec}

According to classic theories of galaxy formation
(e.g., White \& Rees 1978; Mo, Mao, \& White 1998;
see also the comprehensive review by Baugh 2006),
hot-mode spherical accretion dominates the gas accretion history of galaxies.
Once a sufficiently massive dark matter halo has virialized, gas 
collapses through the virial radius of the potential well,
heating via shocks to the virial temperature of the host halo.
As this gas cools over time (largely by collisionally excited line radiation for haloes with $10^4$ K $< T < 10^6$ K, and bremsstrahlung
radiation for haloes with $T \sim 10^7$ K; White 1994) it collapses to form a rotating disk supported by angular momentum which
the cooled gas has been unable to shed.
This gaseous disk gradually grows over time as gas at progressively greater radii is able
to cool and collapse, and is posited to be the home of the bulk of active star formation.

\begin{figure*}
\plotone{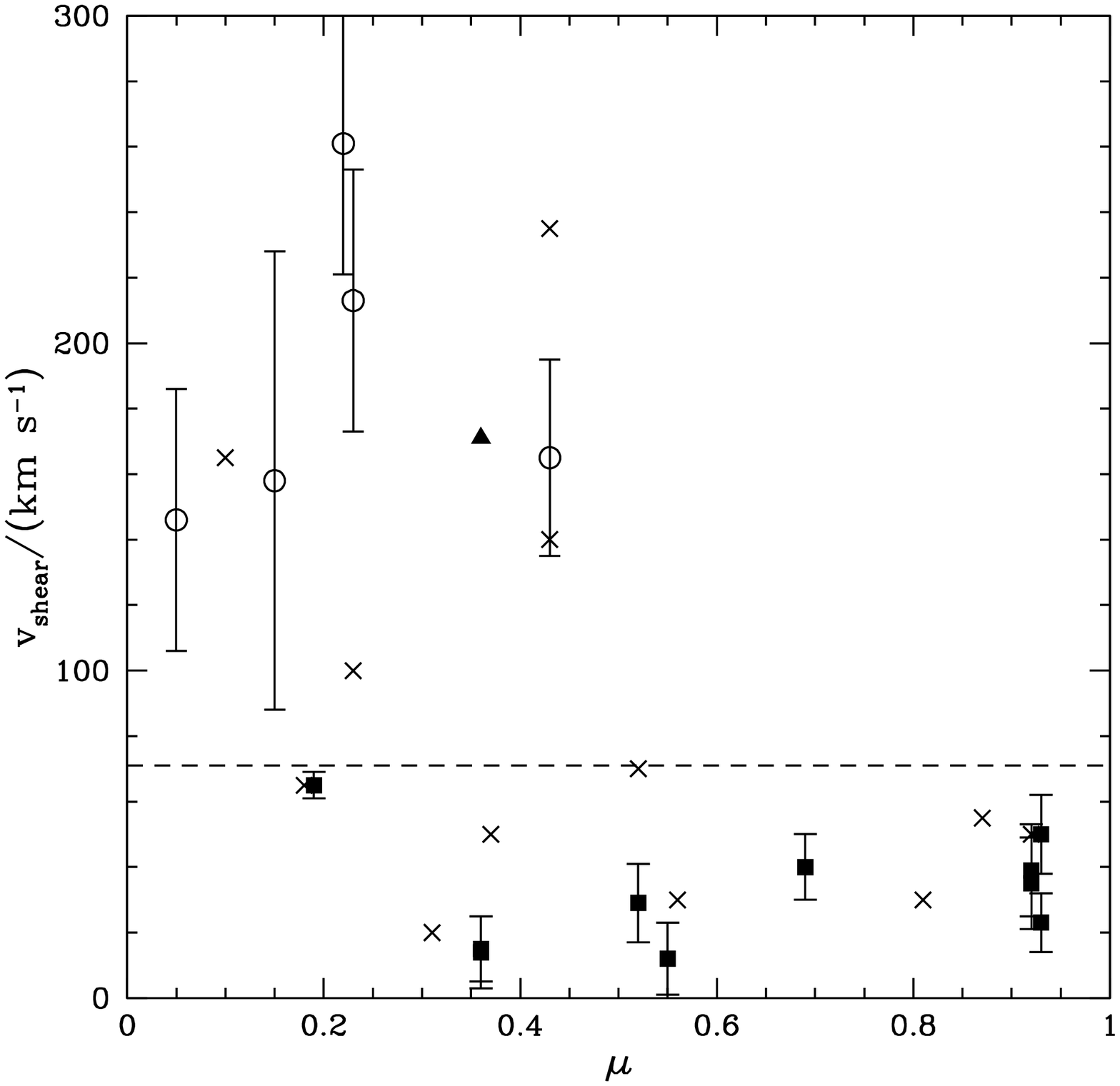}
\caption{Plot of maximum line-of-sight velocity shear versus gas fraction $\mu$.  Symbols are the same as in Figure \ref{mstartrend.fig},
the dashed line denotes $\langle$log$(\sigma_{\rm mean})\rangle$ of our sample.}
\label{mutrend.fig}
\end{figure*}

\begin{figure*}
\plottwo{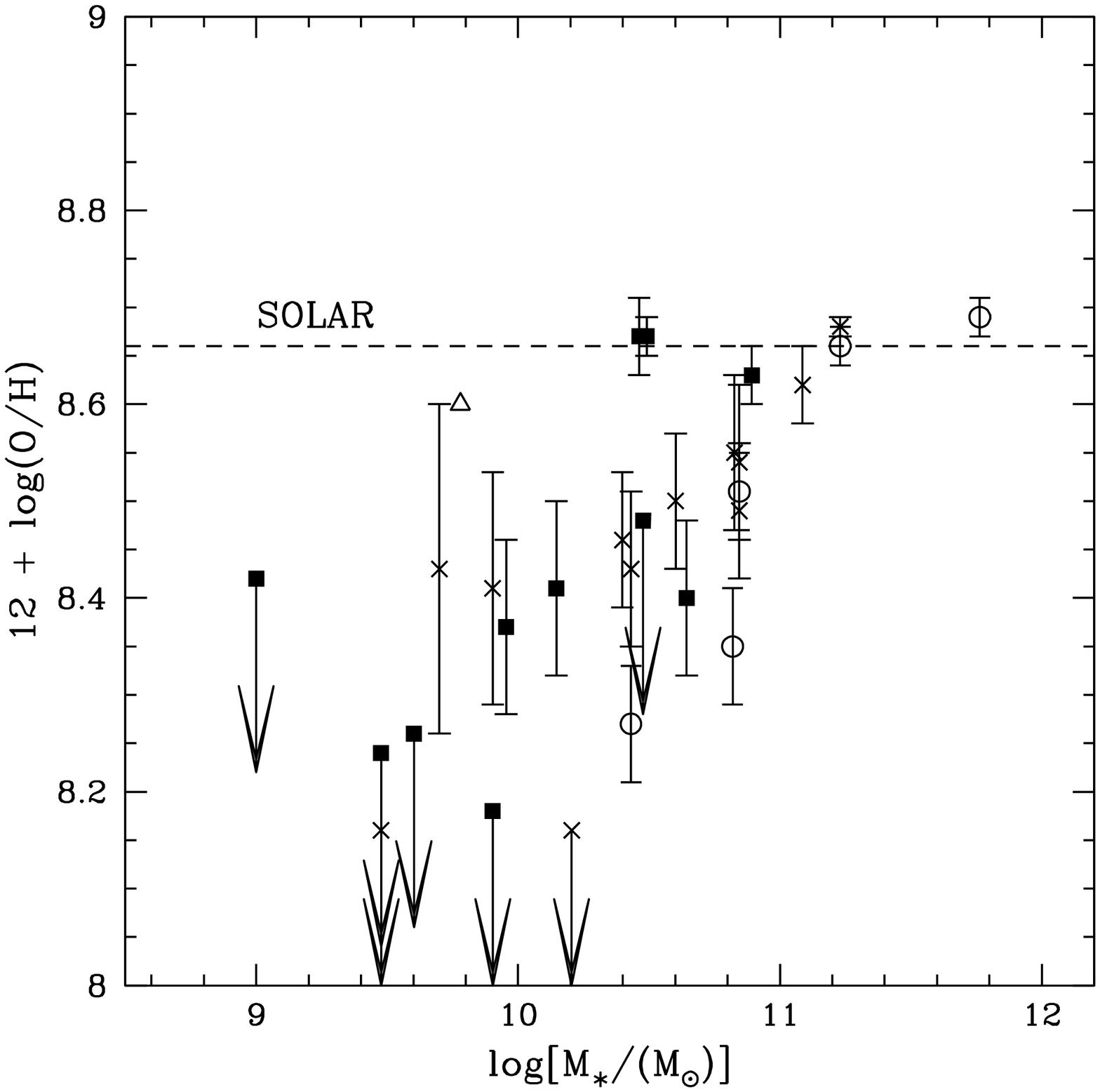}{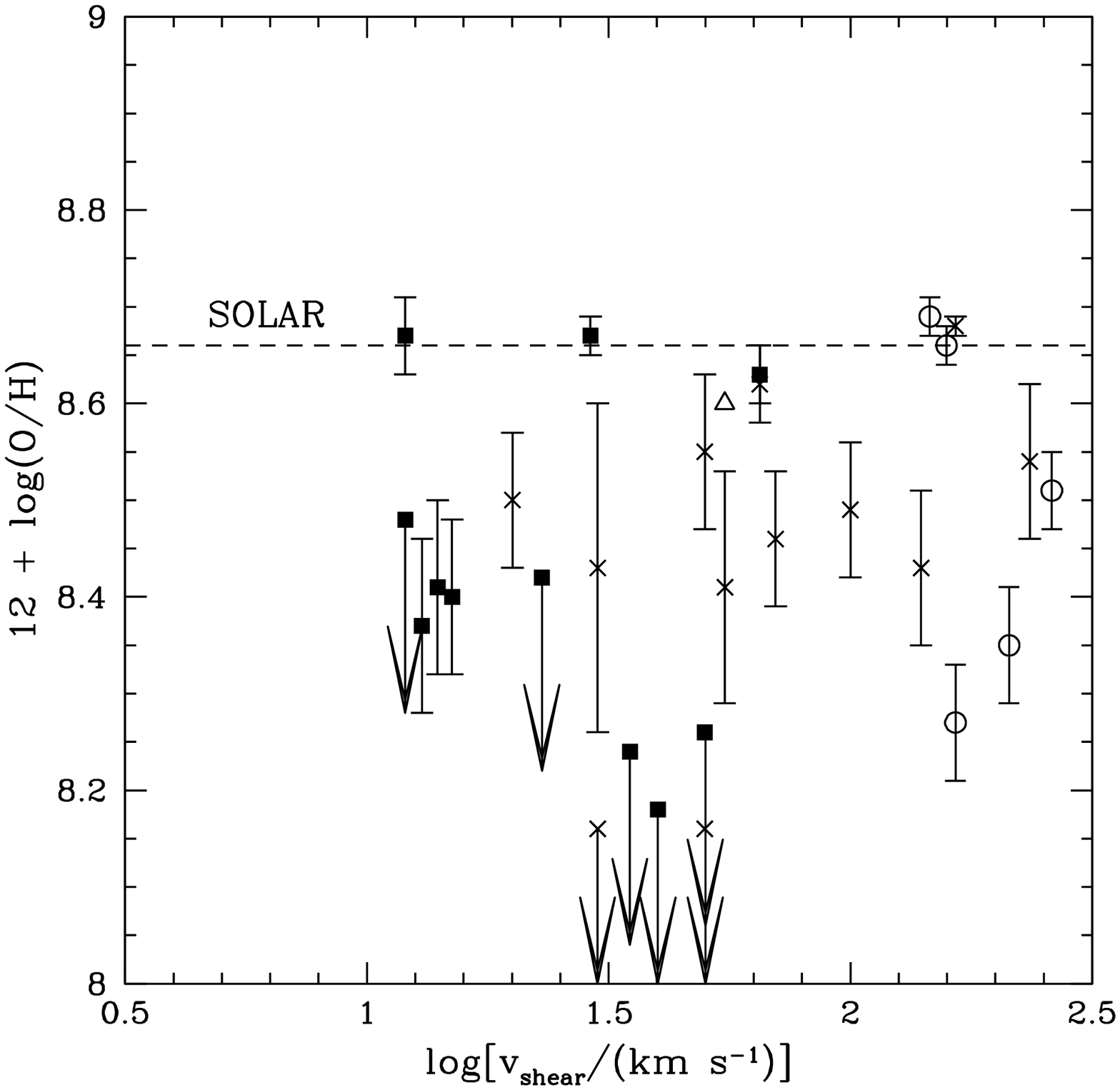}
\caption{Left-hand panel: Relationship between stellar mass and metallicity as traced by the oxygen abundance using the $N2$ calibration of Pettini \& Pagel (2004).
Right-hand panel: Relationship between shear velocity and metallicity.  Symbols are the same as in Figure \ref{mstartrend.fig}, the dashed line indicates solar
metallicity (Asplund et al. 2004).  Upper limits to the metallicity of galaxies undetected in \ntwo are indicated by arrows.}
\label{feh.fig}
\end{figure*}

The rotationally supported gas disks predicted by such a classical model are clearly at odds with the randomly directed kinematics observed
in a substantial fraction of galaxies
at $z \sim 2$, prompting us to revisit our understanding of the complex role of gas accretion in galaxy formation.
Unlike most galaxies in the local universe, the star-forming galaxies at $z \sim 2$ contain extremely large quantities of gas as compared to their stellar mass, 
with the most extreme examples inferred to have gas masses $> 10$ times larger than $M_{\ast}$ (Erb et al 2006c), and comparable to their kinematically-derived dynamical mass.  
Among the UV-selected galaxies in the $H\alpha$ sample of Erb et al. (2006c), the inferred gas fraction $\mu$ (\S \ref{sedmods.sec}) 
decreases with increasing $M_{\ast}$; Figure \ref{mutrend.fig} recasts the trend of shear velocity in terms of $\mu$.
Galaxies with the highest gas fraction $\mu \gtrsim 0.5$ tend to have negligible evidence for rotating kinematic structure and are dominated by randomly oriented velocity
dispersions, while those with the lowest gas fractions $\mu \lesssim 0.5$ have a more obvious preferred axis for their angular momentum with shear
velocities $v_{\rm shear} \gtrsim 100$ \kms.
We note, however, that while we reproduce the stellar mass -- metallicity relation of Erb et al. (2006a; Fig. \ref{feh.fig}, left-hand panel)
there is no clear relationship between oxygen abundance and apparent rotational velocity (Fig. \ref{feh.fig}, right-hand panel).

Given that these galaxies have obviously amassed  large quantities of cold gas within a very small volume
(typically $r \lesssim 1-2$ kpc), the angular momentum of this gas must necessarily be small.
There must exist, therefore, some mechanism that can funnel  gas to small regions within
typical $z \sim 2$ star-forming galaxies efficiently.

Assuming for a moment that a gaseous disk is the first baryonic structure to form in these haloes, gravitational instabilities (e.g. Ostriker \& Peebles 1973) within this gas-dominated disk
may suffice to produce such concentrated gas reservoirs.  
Indeed, Noguchi et al. (1999) showed that efficient cooling mechanisms could lead to fragmentation
of gas-dominated disks into self-gravitating clouds with mass $\sim 10^9 M_{\odot}$, the distribution and relative surface brightness of which
might naturally give rise to an irregular, multiple component morphology 
similar to those observed for  typical star forming galaxies at $z \sim 2$ (e.g., Abraham et al. 1996; Conselice et al. 2005; Elmegreen et al. 2005; Lotz et al. 2006; Law et al. 2007b).
Likewise, simulations by Bournaud et al. (2007) also suggest that ``clump-cluster'' galaxies
(i.e. those with multiple-component morphologies) are consistent with resulting from the fragmentation of unstable primordial disks.
These simulated clumps typically have gas fractions ($\sim$ 50\%), star formation rates ($\sim$ 30 $M_{\odot}$ yr$^{-1}$), and lifetimes ($\sim$ 400 Myr)
comparable to those of the $z \sim 2 - 3$ star forming galaxy population (Erb et al. 2006c).

In gas-rich galaxies such as those of our target sample, these massive gaseous clumps may actually drive the dynamical 
evolution of the entire baryonic component.
As discussed by Bournaud et al. (2007), simulated galaxies during the ``clump-cluster'' phase are highly disturbed with large local velocity 
dispersions $\sim 50 - 80$ \kms (i.e. similar to those
observed in our target galaxies) and negligible coherent rotational signatures on scales $\sim 2 - 3 $ kpc.
At later times ($\sim$ 0.5 - 1 Gyr) these clumps are expected to gradually disperse and sink to the center of the galaxy through dynamical friction, providing a possible
mechanism for bulge formation aided by the massive starburst (SFR $\sim$ 100 $M_{\odot}$ yr$^{-1}$) predicted to occur
when the remaining clouds merge in the galactic center (e.g., Immeli et al. 2004ab; Elmegreen 2008).  At such late times it is likely that a significant stellar population
has been formed which can help stabilize the system against further instability and permit the observation of more regular kinematic structures
(as suggested by Fig. \ref{mstartrend.fig}).


Another possibility however is that the classical gas-disk phase may be bypassed entirely by cold flows of low angular-momentum gas accreted directly from
cosmological filaments (e.g., Kere{\v s} et al. 2005; Dekel \& Birnboim et al. 2006; Birnboim et al. 2007).  
Such flows could be responsible for rapidly delivering large quantities of cold gas to the galaxy's central regions, 
producing physical conditions that in the local universe are achieved only in gas-rich mergers.
The detailed physical structure which might result from such an accretion model is unclear however, 
especially  given the limited ability of numerical models to resolve the relevant
spatial scales.  

Regardless of the mechanism by which this low angular momentum cold gas is acquired, our observational data suggest the possibility that 
many young $z \sim 2$ star-forming galaxies could form roughly spherical early stellar populations in a rapidly-accreted pool of low angular momentum
gas experiencing a rapidly changing gravitational potential dominated by cooling gas.  As the gas fraction of these galaxies decreases this early stellar population
may help stabilize an extended gaseous disk which forms at later times from the gradual, more adiabatic accretion of gas from the galactic halo with increasingly greater angular momentum.
The dynamics of more massive, evolved galaxies might therefore tend, as we observe, to show greater  rotational structure with the formation of such an
extended gaseous disk.  

\begin{figure}[tbp]
\plotone{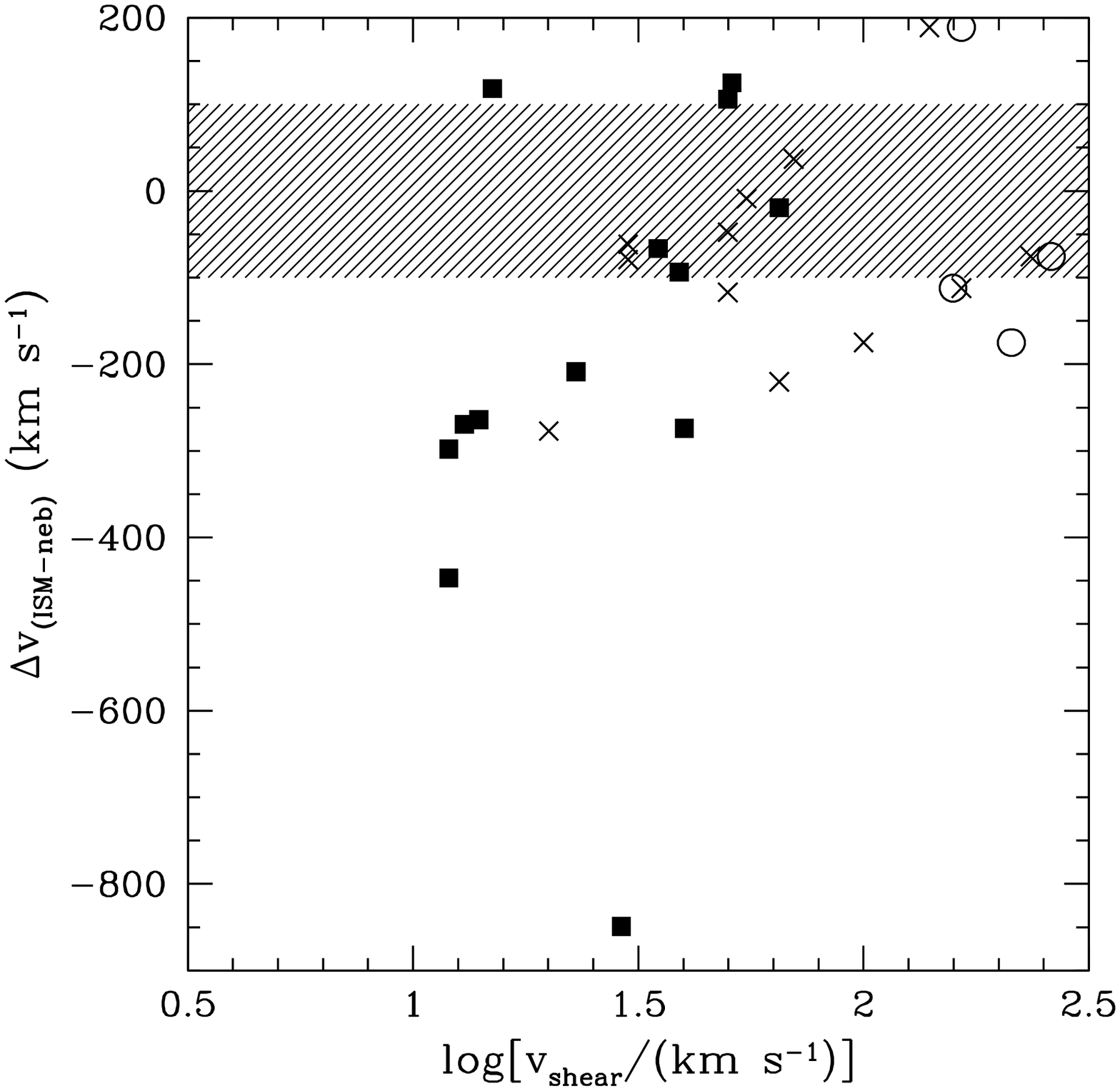}
\caption{Centroid of ISM absorption lines relative to the systemic redshift as a function of $v_{\rm shear}$, symbols are the same as in Figure \ref{mstartrend.fig}.
The hatched region denotes a range within 100 \kms of the systemic redshift (roughly the uncertainty of individual values of  $\Delta v_{\rm (ISM - neb}$).}
\label{uvtrend.fig}
\end{figure}

Such an interpretation may be bolstered by the observation  of a correlation between $v_{\rm shear}$ and
the kinematics of interstellar absorption lines traced by rest-UV spectroscopy (Steidel et al. {\it in prep.}).  
In Figure \ref{uvtrend.fig} we plot the offset $\Delta v_{\rm (ISM - neb)}$ between the absorption and systemic redshifts as a function of $v_{\rm shear}$ for our galaxy sample
and that of F{\"o}rster Schreiber et al. (2006) for which we have rest-UV spectra.
While the majority of $z \sim 2$ galaxies have rest-UV absorption features 
with centroids  blueshifted relative to the systemic
velocity by $\sim 200$ \kms (e.g. Pettini et al. 2002; Shapley et al. 2003; Steidel et al. {\it in prep.}), 
there is a tendency for galaxies with the greatest observed velocity shear to have lower $\Delta v_{\rm (ISM - neb)}$, indicating increased optical depth to UV photons at low
systemic velocities (as noted previously by Erb et al. 2006a and Law et al. 2007a).
Although the properties of the outflowing gas traced by interstellar absorption lines are generally unrelated to the kinematics of the nebular emission lines
(as any nebular emission within the outflowing gas is  too low surface brightness to be detectable) it is possible that galaxies with a more well-defined kinematic axis
could contain a greater fraction of their ISM at close to the systemic redshift (see discussion by Steidel et al. {\it in prep}).

This relationship between gas content and $v_{\rm shear}$ may begin to break down however at lower redshift.
Observations by Wright et al. (2009) at $z \sim 1.6$ find that some galaxies in a similar range of stellar masses and gas fractions to our own sample 
(and selected in a similar manner) exhibit significant kinematic gradients greater than their
apparent velocity dispersion.  The applicability of this relation at different redshifts and halo masses is therefore uncertain, but might naturally be expected to evolve along
with changes in the cooling time of gas in galactic haloes.

\section{SUMMARY}
\label{summary.sec}

We have presented Keck-LGSAO observations of 13 star-forming galaxies at redshift $z \sim 2 - 3$
which are representative of the typical stellar mass of the optically-selected galaxy sample that dominates the global star formation activity at these epochs.
This population of galaxies has a gas-phase velocity dispersion $\sigma \sim 70$ \kms
that cannot be explained by beam-smearing effects and is an intrinsic property of the gas.
These random motions are larger than the amplitude of any systemic velocity shear $v_{\rm shear}$ within the central 2-3 kpc, although in some cases
$v_{\rm shear}$ may increase if the observed velocity gradient persists to larger radii and fainter surface brightness.

The dominance of random motions (or the absence of significant shear velocity) appears to be related to the stellar mass and inferred gas fraction: 
galaxies with larger velocity gradients tend to have larger stellar masses ($2 - 20$ times larger than the mean for the parent sample), with smaller values of the gas fraction
($\mu < 0.5$).
While there is no similar relation between metallicity and kinematics, there is some evidence to suggest that galaxies with the greatest large-scale velocity structure 
may contain a greater fraction of their total interstellar medium close to the systemic redshift.

Although the observed kinematics of the high redshift galaxies are similar to those of merger-driven starburst galaxies in the local universe, it is unlikely that 
mergers are the primary reason for the observed dispersion-dominated kinematics at $z \sim 2 - 3$.  Instead, more efficient cooling and settling of larger 
amounts of low angular momentum gas (whether attributed to ``cold accretion'' or some other process) may generically give high redshift galaxies properties 
that are produced only in mergers in the local universe.  As the large, centrally concentrated gas supply is rapidly converted into stars, the growing stellar 
population may help stabilize the galaxy and permit the formation of stable disk-like structures from higher angular momentum gas accreted over longer timescales.

While such a qualitative picture remains  speculative given the current state of both observations and theory, it is without doubt an over-simplification to classify 
high redshift galaxies as either ``disks'' or ``mergers'' ; in fact, many, and possibly even most, are neither.  High gas-phase velocity dispersions (and small velocity gradients) 
are evidently a natural consequence of the instablities resulting when cold gas becomes dynamically dominant, as it seems to be in the central few kpc of a large fraction of 
star-forming galaxies at $z \sim 2-3$.  Learning more precisely how and why this occurs may be the key to understanding galaxy formation.

\acknowledgements

The authors thank Randy Campbell, Al Conrad, and Jim Lyke for their invaluable assistance obtaining the observations presented herein.
DRL also thanks Andrew Benson and Mark Swinbank for constructive discussions, Naveen Reddy for providing stellar mass function data,
and the anonymous referee for helpful suggestions which improved the final draft of this manuscript.
DRL and CCS have been supported by grants AST-0606912 and AST-0307263 from the US National Science Foundation.
Additional support for this work was provided by NASA through Hubble Fellowship grant \# HF-01221.01
awarded by the Space Telescope Science Institute, which is operated by the Association of Universities for Research in Astronomy, Inc., for NASA, under contract NAS 5-26555.
Finally, we wish to extend thanks to those of Hawaiian ancestry on whose sacred mountain we are privileged to be guests.

\setcounter{section}{0}
\renewcommand{\thesection}{A.\arabic{section}}
\section{QUANTIFYING THE MEAN PROPERTIES OF INCLINED SYSTEMS}
\label{append.sec}

In cases for which there is obvious morphological or kinematic symmetry indicative of an inclined system, foreshortening of the respective contours
may be used to correct apparent rotational velocities for the inclination with confidence.  In other cases however the degree of such foreshortening may not be obvious, 
and it is frequently popular to apply an
 `average' inclination correction to determine the intrinsic rotational velocity in a statistical sense.  Confusingly, the value of this average correction frequently differs in the literature,
with some authors quoting $\langle i \rangle = 45^{\circ}$ and $\langle \sin i \rangle = \frac{2}{\pi} = 0.64$ (e.g., Erb et al. 2006c; Bouch{\'e} et al. 2007) and others adopting
$\langle i \rangle = 57.3^{\circ}$ and $\langle \sin i \rangle = \frac{\pi}{4} = 0.79$ (e.g. Rix et al. 1997;
Weiner et al. 2006).  In an effort to foster consistency we present a simple derivation of the correct factor instead of simply stating an adopted value.

The easiest way of visualizing the problem is to imagine the range of possible inclinations $i$ at which a disk can be viewed.  While $i$ ranges from 
0$^{\circ}$ to 90$^{\circ}$, the average value of $i$ is not 45$^{\circ}$, since there are many more ways in which a disk may be viewed from its edge than from its pole.  Explicitly, we
frame the problem as follows.
Given a collection of disks oriented isotropically in space, we wish to determine the mean expected inclination $i$ between our line of sight to a given disk
and the vector normal to the disk (where $i=0^{\circ}$ represents a disk viewed face-on).  It is simplest to reduce this problem to the case of a single disk
and consider it from the reference frame of the disk itself.
Assuming that a viewing location
is chosen at random, the distribution of possible lines of sight will uniformly cover the sky as seen from the reference frame of the disk.
The probability $dP$ that a viewer is located within a given patch of sky is therefore proportional to the differential solid angle $d\Omega$ subtended by that
patch of sky.
Adopting a spherical polar coordinate system $\theta, \phi$ centered on the disk where $\phi$ is the azimuthal angle and $\theta$ (ranging from $-\frac{\pi}{2}$ to $\frac{\pi}{2}$)
is the polar angle the differential solid angle is given by
\begin{equation}
d\Omega = \cos \theta \, d\theta \, d\phi .
\end{equation}

It is therefore possible to calculate the expectation value $\langle i \rangle$ by integrating over the entire sky.
Noting the symmetry present in the problem, we may omit the integral over the azimuthal coordinate $\phi$ and collapse the polar integral to consider only the range
$\theta = 0$ to $\frac{\pi}{2}$, reducing the differential solid angle to $d\Omega = \cos \theta \, d\theta$.  
In this range, the inclination is related to the polar angle by $\theta = \frac{\pi}{2} - i$.
Therefore,
\begin{equation}
\langle i \rangle  = \frac{\int_0^{\frac{\pi}{2}}i \, \cos \theta \, d\theta}{\int_0^{\frac{\pi}{2}} \cos \theta \, d\theta} .
\end{equation}

Substituting the identities $\cos \theta = \sin i$ and $d\theta = -di$ we obtain
\begin{equation}
\langle i \rangle = \frac{\int_{0}^{\frac{\pi}{2}} i \, \sin i \, di}{\int_{0}^{\frac{\pi}{2}} \sin i \, di} = \frac{[\sin i - i \, \cos i]_0^{\frac{\pi}{2}}}{[- \cos i]_0^{\frac{\pi}{2}}} = 57.3^{\circ} .
\end{equation}

Since the observed velocities of an inclined disk are effectively ``foreshortened'' by a factor sin $i$, it is usually of interest to calculate the mean velocity reduction factor.
Similarly to the calculations performed above, we find
\begin{equation}
\langle \sin i \rangle = \frac{\int_0^{\frac{\pi}{2}} \sin i \, \cos \theta \, d\theta}{\int_0^{\frac{\pi}{2}} \cos \theta \, d\theta} = \frac{\int_0^{\frac{\pi}{2}} \sin ^2 i \, di}{\int_0^{\frac{\pi}{2}} \sin i \, di} = \frac{\pi}{4} = 0.79 .
\end{equation}

\renewcommand{\thesection}{\thechapter.\arabic{section}}
\renewcommand{\theequation}{\thechapter.\arabic{equation}}

\clearpage


\begin{deluxetable}{lccccccccc}
\tablecolumns{10}
\tablewidth{0pc}
\tabletypesize{\scriptsize}
\tablecaption{Observing Details}
\tablehead{
\colhead{} & \colhead{} & \colhead{R.A.} & \colhead{Decl.} & \colhead{Observing} & \colhead{Exposure Time\tablenotemark{b}} & \colhead{} & \colhead{Em.} & \colhead{Scale} & \colhead{$\theta_{\rm PSF}$\tablenotemark{d}}\\
\colhead{Galaxy} & \colhead{$z_{\rm neb}$\tablenotemark{a}} & \colhead{(J2000.0)} & \colhead{(J2000.0)} & \colhead{Run} & \colhead{(seconds)} & \colhead{Filter} & \colhead{Line\tablenotemark{c}} & \colhead{(mas)} & \colhead{(mas)}}
\startdata
\multicolumn{10}{c}{Detections}\\
\hline
Q0449-BX93 & 2.0067 & 04:52:15.417 & $-$16:40:56.88 & 2006 Oct & 16200 & Kn1 & \Ha & 50 & 115/170\\
Q1217-BX95 & 2.4244 & 12:19:28.281 & $+$49:41:25.90 & 2008 Jun & 6300 & Kn4 & \Ha & 50 & 95/135\\
HDF-BX1564 & 2.2228 & 12:37:23.470 & $+$62:17:20.00 & 2007 Jun & 3600 & Kn2 & \Ha & 50 & 150/290\\
Q1623-BX453 & 2.1820 & 16:25:50.854 & $+$26:49:31.28 & 2006 Jun & 9000 & Kn2 & \Ha & 50 & 70/140\\
Q1623-BX502 & 2.1557 & 16:25:54.385 & $+$26:44:09.30 & 2007 Jun & 9900\tablenotemark{e} & Kn2 & \Ha & 50 & 195/220\\
Q1623-BX543 & 2.5211 & 16:25:57.707 & $+$26:50:08.60 & 2008 Jun & 11700 & Hn5 & \othree & 50 & 95/145\\
Q1700-BX490 & 2.3958 & 17:01:14.830 & $+$64:09:51.69 & 2008 Jun & 11700 & Kn4 & \Ha & 50 & 75/125\\
Q1700-BX710 & 2.2947 & 17:01:22.128 & $+$64:12:19.21 & 2006 Jun & 5400 & Kn3 & \Ha & 50 & 70/140\\
Q1700-BX763 & 2.2919 & 17:01:31.463 & $+$64:12:57.67 & 2008 Jun & 12600 & Kn3 & \Ha & 50 & 75/190\\
DSF2237a-C2 & 3.3172 & 22:40:08.298 & $+$11:49:04.89 & 2006 Jun & 5400 & Kn3 & \othree & 50 & 70/140\\
Q2343-BX418 & 2.3053 & 23:46:18.582 & $+$12:47:47.77 & 2008 Jun & 6300 & Kn3 & \Ha & 50 & 100/150\\
Q2343-BX513 & 2.1082 & 23:46:11.133 & $+$12:48:32.54 & 2006 Oct & 12600 & Kn1 & \Ha & 50 & 110/165\\
Q2343-BX660 & 2.1739 & 23:46:29.447 & $+$12:49:45.93 & 2008 Sep & 10800 & Kn2 & \Ha & 50 & 90/140\\
\hline
\multicolumn{10}{c}{Non-Detections}\\
\hline
Q0100-BX210 & 2.279\tablenotemark{f} & 01:03:11.996 & $+$13:16:18.32 & 2006 Oct & 1800\tablenotemark{e} & Kn3 & \Ha & 50 & 120\\
 &  &  &  & 2007 Sep & 3600\tablenotemark{e} & Hn3 & \othree & 50 & 90\\
 &  &  &  & 2007 Sep & 5400\tablenotemark{e} & Kn3 & \Ha & 50 & 120\\
HDF-BX1311 & 2.4843\tablenotemark{f} & 12:36:30.514 & $+$62:16:26.00 & 2007 Jun & 4500\tablenotemark{e} & Kn4 & \Ha & 50 & 140\\
HDF-BX1439 & 2.1865\tablenotemark{f} & 12:36:53.660 & $+$62:17:24.00 & 2008 Jun & 2700 & Kn2 & \Ha & 50 & 70\\
 &  &  &  & 2008 Jun & 7200\tablenotemark{e} & Kn2 & \Ha & 100 & 80\\
Q1623-BX455\tablenotemark{i} & 2.4079 & 16:25:51.664 & $+$26:46:54.60 & 2008 Jun & 5400 & Kn4 & \Ha & 50 & 85\\
 &  &  &  & 2007 Sep & 3600 & Hn4 & \othree & 50 & 70\\
Q1623-BX663 & 2.4333\tablenotemark{f} & 16:26:04.586 & $+$26:47:59.80 & 2007 Jun & 5400\tablenotemark{e} & Kn4 & \Ha & 50 & 100\\
Q1700-BX563 & 2.292\tablenotemark{g} & 17:01:15.875 & $+$64:10:26.15 & 2007 Jun & 1800\tablenotemark{e} & Kn3 & \Ha & 50 & 210\\
Q1700-BX691 & 2.1895\tablenotemark{f} & 17:01:06.117 & $+$64:12:09.70 & 2006 Jun & 10800\tablenotemark{h} & Kn2 & \Ha & 50 & ...\\
 &  &  &  & 2007 Jun & 5400\tablenotemark{e} & Kn2 & \Ha & 50 & 200\\
Q2206-BX102 & 2.2104\tablenotemark{f} & 22:08:50.751 & $-$19:44:08.24 & 2007 Sep & 5400 & Kn2 & \Ha & 50 & 90\\
Q2343-BX389 & 2.1716\tablenotemark{f} & 23:46:28.911 & $+$12:47:33.90 & 2007 Jun & 8100\tablenotemark{e} & Kn2 & \Ha & 50 & 160\\
Q2343-BX442 & 2.1760\tablenotemark{f} & 23:46:19.362 & $+$12:48:00.10 & 2007 Sep & 9000\tablenotemark{e} & Kn2 & \Ha & 50 & 140\\
Q2343-BX587\tablenotemark{i} & 2.2429 & 23:46:29.192 & $+$12:49:03.71 & 2006 Oct & 5400 & Kn3 & \Ha & 50 & 105\\
\enddata
\tablenotetext{a}{Vacuum heliocentric redshift of primary nebular emission line.}
\tablenotetext{b}{Total observing time, mean value for detected sources was $\sim 2$ hrs.}
\tablenotetext{c}{Primary targeted emission line.}
\tablenotetext{d}{FWHM of the $K$-band PSF (mas) during on-axis TT star observation (before/after spatial smoothing respectively).}
\tablenotetext{e}{Poor observing conditions.}
\tablenotetext{f}{Redshifts estimated from NIRSPEC spectra.}
\tablenotetext{g}{Redshift estimated from rest-UV spectrum.}
\tablenotetext{h}{Individual exposures were each 300 seconds.}
\tablenotetext{i}{Galaxy detected, but quality too poor for analysis.}
\label{targets.table}
\end{deluxetable}

\clearpage


\begin{deluxetable}{lcccccccccc}
\tablecolumns{11}
\tablewidth{0pc}
\tabletypesize{\scriptsize}
\tablecaption{Nebular Line Fluxes}
\tablehead{
\colhead{} & 
\colhead{$\lambda_{\rm neb}$\tablenotemark{a}} & 
\colhead{} & 
\colhead{} & 
\colhead{} &
\colhead{$F_{\othree}$\tablenotemark{c}} & 
\colhead{$L_{\rm neb}$\tablenotemark{e}} & 
\colhead{$F_{\othree}$\tablenotemark{c}} & 
\colhead{$F_{\ntwo}$\tablenotemark{c}} & 
\colhead{$F_{\ntwo}$\tablenotemark{c}} & \colhead{}\\
\colhead{Galaxy} & 
\colhead{(\AA)} & 
\colhead{$z_{\rm neb}$\tablenotemark{b}} & 
\colhead{$F_{\Ha}$\tablenotemark{c}} & 
\colhead{$F_{\rm DKE}$\tablenotemark{d}} &
\colhead{($\lambda5007$)} & 
\colhead{($10^{42}$ erg s$^{-1}$)} & 
\colhead{($\lambda4960$)} & 
\colhead{($\lambda6549$)} & 
\colhead{($\lambda6585$)} &
\colhead{$12 +$ log($O/H$)\tablenotemark{f}}
}
\startdata
Q0449-BX93 & 19737.8 & 2.0067 & $6.8\pm0.2$ & ... & ... & $3.0\pm0.1$ & ... & $1.3\pm0.4$ & $0.8\pm0.3$ & $8.37\pm0.09$ \\

Q1217-BX95 & 22479.6 & 2.4244 & $6.5\pm0.4$ & ... & ... & $6.4\pm0.4$ & ... & $\leq 1.2$ & $\leq 1.2$ & $\leq 8.48$\\

HDF-BX1564 & 21156.7 & 2.2228 & $9.7\pm0.8$ & $17.2\pm1.4$& ... & $7.0\pm0.4$ & ... & $\leq 1.5$ & $3.9\pm0.5$ & $8.67\pm0.04$ \\

Q1623-BX453 & 20888.3 & 2.1820 & $16.4\pm0.4$ & $27.6\pm0.4$ & ... & $12.8\pm0.3$ & ... & $3.0\pm0.4$ & $6.5\pm0.4$ & $8.67\pm0.02$ \\

Q1623-BX502 & 20715.8 & 2.1557 & $10.0\pm0.2$ & $26.4\pm0.8$& ... & $5.3\pm0.1$ & ... & $\leq 0.7$ & $\leq 0.7$ & $\leq 8.24$ \\

Q1623-BX543 & 17634.4 & 2.5211 & ... & $17.2\pm1.4$ & $25.0\pm0.5$ & $42.5\pm0.8$ & $8.0\pm0.5$ & ... & ... & ... \\

Q1700-BX490 & 22291.8 & 2.3958 & $31.6\pm0.8$ & $35.4\pm1.2$& ... & $34.6\pm0.9$ & ... & $\leq 2.4$ & $ \leq 2.4$ & $\leq 8.26$\\

Q1700-BX710 & 21628.4 & 2.2947 & $4.5\pm0.2$ & ... & ... & $3.4\pm0.2$ & ... & $\leq 0.6$ & $0.6\pm0.2$ & $8.40\pm0.08$  \\

Q1700-BX763 & 21609.8 & 2.2919 & $2.2\pm0.3$ & ... & ... & $1.3\pm0.2$ & ... & $\leq 0.9$ & $0.3\pm0.3$ & $8.41\pm0.25$ \\

DSF2237a-C2 & 21621.8 & 3.3172 & ... & ... & $7.9\pm 0.4$ & $16.8\pm0.9$ & $2.0\pm0.3$ & ... & ... & ... \\

Q2343-BX418 & 21698.2 & 2.3053 & $8.2\pm0.4$ & $16.0\pm0.4$ & ... & $3.8\pm0.2$ & ... & $\leq 1.2$ & $\leq 1.2$ & $\leq 8.42$ \\

Q2343-BX513 & 20404.3 & 2.1082 & $5.4\pm0.2$ & $20.2\pm0.8$ & ... & $2.9\pm0.1$ & ... & $\leq 0.6$ & $1.8\pm0.2$ & $8.63\pm0.03$ \\

Q2343-BX660 & 20835.4 & 2.1739 & $11.2\pm0.2$ & $18.8\pm0.8$ & ... & $4.2\pm0.1$ & ... & $\leq 0.6$ & $\leq 0.6$ & $\leq 8.18$ \\
\enddata
\label{fluxes.table}
\tablenotetext{a}{Vacuum heliocentric wavelength of primary nebular emission: \othree $\,$ $\lambda$5007 for DSF2237a-C2 and Q2343-BX415, \ntwo $\,$ for Q1623-BX455, \Ha $\,$  for all others.}
\tablenotetext{b}{Heliocentric redshift of primary nebular emission line.}
\tablenotetext{c}{Emission line flux in units of $10^{-17}$ erg s$^{-1}$ cm$^{-2}$.  Uncertainties quoted are $1\sigma$ and based on random errors,
global systematic uncertainty is $\sim 30$\%.  Limits represent 3 $\sigma$ limits.}
\tablenotetext{d}{\Ha\ emission line flux observed by Erb et al. 2006c (includes a factor of 2 correction for aperture losses).}
\tablenotetext{e}{Extinction-corrected primary nebular emission line luminosity (\Ha $\,$ or \othree).  Uncertainties quoted are based on random errors,
global systematic uncertainty is $\sim 30$\%.}
\tablenotetext{f}{Oxygen abundance using the N2 calibration from Pettini \& Pagel (2004).}
\end{deluxetable}

\clearpage


\begin{deluxetable}{lcccccc}
\tablecolumns{7}
\tablewidth{0pc}
\tabletypesize{\scriptsize}
\tablecaption{OSIRIS Morphologies}
\tablehead{
\colhead{} & \colhead{$I$\tablenotemark{a}} & \colhead{$r$\tablenotemark{b}} & \colhead{$d_{\rm 2c}$\tablenotemark{c}} &
\colhead{} & \colhead{} & \colhead{} 
\\
\colhead{Galaxy} & \colhead{(kpc$^2$)} & \colhead{(kpc)} & \colhead{(kpc)} &
\colhead{$G$\tablenotemark{d}} & \colhead{$\Psi$\tablenotemark{e}} & \colhead{$M_{20}$} 
}
\startdata
Q0449-BX93 & 5.0 $\pm$ 0.8 & 1.3 $\pm$ 0.1 & ... & 0.24 & 1.2 & $-$1.08 \\

Q1217-BX95 & $1.0\pm0.4$ & $0.6\pm0.1$ & ... & 0.13 & 0.8 & $-$0.89 \\

HDF-BX1564 & $5.3\pm2.1$ (NW) & $1.3\pm0.2$ (NW) & 7.0 & 0.10 & 13.6 & $-$0.51 \\
 & $\leq 3$ (SE) & $\leq 1$ (SE) & &\\

Q1623-BX453 & 6.1 $\pm$ 0.5 & 1.4 $\pm$ 0.1 & ... & 0.22 & 1.3 & $-$1.36 \\

Q1623-BX502 & $6.2\pm1.3$ & $1.4\pm0.2$ & ... & 0.21 & 0.7 & $-$1.48 \\

Q1623-BX543 & $3.6\pm0.5$ (N) & $1.1\pm0.1$ (N) & 6.7 & 0.19 & 9.7 & $-$1.09 \\
 & $1.5\pm0.5$ (S) & $0.7\pm0.1$ (S) & &\\

Q1700-BX490 & $8.3\pm0.4$ (W) & $1.6\pm0.1$ (W) & 3.4 & 0.15 & 5.8 & $-$1.40 \\
 & $1.6\pm0.4$ (E) & $0.7\pm0.1$ (E) & &\\

Q1700-BX710 & $4.1\pm0.5$ & $1.1\pm0.1$ & ... & 0.23 & 0.5 & $-$1.51 \\

Q1700-BX763 & $3.9\pm0.9$ & $1.1\pm0.1$ & ... & 0.14 & 1.5 & $-$1.22 \\

DSF2237a-C2 & 2.7 $\pm$ 0.4 & 0.9 $\pm$ 0.1 & ... & 0.17 & 2.4 & $-$1.16 \\

Q2343-BX418 & $2.2\pm0.6$ & $0.8\pm0.1$ & ... & 0.13 & 1.8 & $-$1.32 \\

Q2343-BX513 & $6.2\pm0.7$ & $1.4\pm0.1$ & ... & 0.19 & 1.1 & $-$1.28 \\

Q2343-BX660 & $8.5 \pm 0.5$ & $1.6 \pm 0.1$ & ... & 0.18 & 1.7 & $-$0.99 \\
\enddata
\label{morphs.table}
\tablenotetext{a}{Area of nebular emission.  Uncertainty represents half the PSF correction.  Individual components are identified in brackets.}
\tablenotetext{b}{Radius of nebular emission.  Uncertainty represents half the PSF correction.  Individual components are identified in brackets.}
\tablenotetext{c}{Distance between the two morphological components.}
\tablenotetext{d}{Gini.}
\tablenotetext{e}{Multiplicity.}
\end{deluxetable}

\clearpage


\begin{deluxetable}{lcccccccc}
\tablecolumns{9}
\tablewidth{0pc}
\tabletypesize{\scriptsize}
\tablecaption{Kinematic Properties}
\tablehead{
\colhead{} & \colhead{$\sigma_{\rm mean}$\tablenotemark{a}} & \colhead{$\sigma_{\rm net}$\tablenotemark{b}} & \colhead{$v_{\rm shear}$\tablenotemark{c}} 
& \colhead{$v_{\rm 2c}$\tablenotemark{d}}
& \colhead{$\Delta v_{\rm (ISM - neb)}$\tablenotemark{e}}
& \colhead{PA\tablenotemark{f}} & \colhead{} & \colhead{$M_{\rm dyn}$\tablenotemark{g}}\\
\colhead{Galaxy} & \colhead{(km s$^{-1}$)} & \colhead{(km s$^{-1}$)} & \colhead{(km s$^{-1}$)} & \colhead{(km s$^{-1}$)} & \colhead{(km s$^{-1}$)} 
& \colhead{($^{\circ}$)} & \colhead{$v_{\rm shear}/\sigma_{\rm mean}$} 
&\colhead{($10^{10} M_{\odot}$)}}
\startdata
Q0449-BX93 & $56\pm20$ & $71\pm5$ & $13\pm8$  & ... & $-270$ & 330 & $0.2\pm0.2$ & 0.5\\
Q1217-BX95 & $69\pm23$ & $61\pm5$ & $12\pm13$  & ... & $-298$ & 110 & $0.2\pm0.2$ & 0.3\\
HDF-BX1564 & $60\pm13$(NW) & $103\pm14$ & $12\pm12$ (NW)  & 171 & $-447$ & 325 (NW) & $0.2\pm0.2$ (NW) & 0.5 (NW)\\
           & $86\pm8$(SE) &  & $5\pm12$ (SE) & &  & 325 (SE) & $0.1\pm0.1$ (SE) & $\leq 0.9$ (NW)\\
Q1623-BX453 & $78\pm23$ & $94\pm5$ & $29\pm10$  & ... & $-849$ & 45 & $0.4\pm0.2$ & 1.0\\
Q1623-BX502 & $67\pm20$ & $68\pm3$ & $35\pm11$  & ... & $-67$ & 30 & $0.5\pm0.3$ & 0.7\\
Q1623-BX543 & $139\pm32$ (N) & $153\pm7$ & $39\pm4$ (N) & 125  & $-94$ & 315 (N) & $0.3\pm0.1$ (N) & 2.5 (N)\\
            & $60\pm10$ (S) & & $22\pm9$ (S) & & & 225 (S) & $0.4\pm0.2$ (S) & 0.3 (S)\\
Q1700-BX490 & $89\pm25$ (W) & $122\pm6$ & $50\pm10$ (W)  & 109 & 106 & 200 (W) & $0.6\pm0.2$ (W) & 1.5 (W)\\
            & $65\pm29$ (E) & & $22\pm9$ (E) & & & 270 (E) & $0.5\pm0.3$ (E) & 0.3 (E)\\
Q1700-BX710 & $68\pm25$ & $67\pm7$ & $15\pm16$  & ... & 118 & 0 & $0.2\pm0.3$ & 0.6\\
Q1700-BX763 & $50\pm14$ & $53\pm5$ & $14\pm16$  & ... & $-264$ & 320 & $0.3\pm0.3$ & 0.3\\
DSF2237a-C2 & $89\pm20$  & $99\pm10$ & $51\pm14$  & ... & 125 & 220 & $0.6\pm0.2$ & 0.8\\
Q2343-BX418 & $61\pm17$ & $70\pm5$ & $23\pm12$  & ... & $-209$ & 70 & $0.4\pm0.2$ & 0.3\\
Q2343-BX513 & $100\pm25$ & $102\pm6$ & $65\pm16$  & ... & $-19$ & 180 & $0.7\pm0.2$ & 1.6\\
Q2343-BX660 & $71\pm16$ & $73\pm$  & $40\pm9$  & ... & $-274$ & 45 & $0.6\pm0.2$ & 0.9\\
\enddata
\label{kinematics.table}
\tablenotetext{a}{Flux-weighted mean and standard deviation of velocity dispersions in individual spaxels in the OSIRIS map.}
\tablenotetext{b}{Velocity dispersion of the spatially collapsed object spectrum.  Uncertainty quoted is based on Monte Carlo analysis of synthetic spectra.}
\tablenotetext{c}{Shear velocity $v_{\rm shear} = \frac{1}{2} (v_{\rm max} - v_{\rm min})$.}
\tablenotetext{d}{Velocity difference between two discrete components.}
\tablenotetext{e}{Kinematic offset between ISM absorption line redshift and nebular emission line redshift.}
\tablenotetext{f}{Position angle of steepest velocity gradient.}
\tablenotetext{g}{Single-component dynamical mass within the radius probed by nebular emission.}
\end{deluxetable}

\clearpage


\begin{deluxetable}{lccccccccc}
\tablecolumns{10}
\tablewidth{0pc}
\tabletypesize{\scriptsize}
\tablecaption{Stellar Population Parameters}
\tablehead{
\colhead{} & \colhead{} & \colhead{$M_{\ast}$\tablenotemark{b}} & \colhead{Age\tablenotemark{c}} & \colhead{SFR$_{\rm SED}$\tablenotemark{d}} & \colhead{SFR$_{\rm neb}$\tablenotemark{e}} 
& \colhead{SFR$_{\rm DKE}$\tablenotemark{f}} & \colhead{} &
\colhead{$M_{\rm gas}$\tablenotemark{h}} & \colhead{}\\
\colhead{Galaxy} & \colhead{$K_{\rm s}$\tablenotemark{a}} & \colhead{($10^{10} M_{\odot}$)} & \colhead{(Myr)} & \colhead{($M_{\odot}$ yr$^{-1}$)}  & \colhead{($M_{\odot}$ yr$^{-1}$)} & \colhead{($M_{\odot}$ yr$^{-1}$)} & \colhead{$E(B-V)$\tablenotemark{g}} &
\colhead{($10^{10} M_{\odot}$)} & \colhead{$\mu$\tablenotemark{i}}}
\startdata
\multicolumn{10}{c}{Detections}\\
\hline
Q0449-BX93 & 20.51 [20.59] & 0.9 & 203 & 47 & 13 & ... & 0.135 & ... & ...\\
Q1217-BX95 & 20.69 [20.79] & 3.0 & 571 & 52 & 28  & ... & 0.235 & ... & ...\\
HDF-BX1564 & 19.62 [19.70] & 2.9 & 571 & 53 & 31 & 34 &  0.205 & 3.6 & 0.55 \\
Q1623-BX453 & 19.76 [20.00] & 3.1 & 404 & 77 & 57 & 100 &  0.245 & 3.3 & 0.52 \\
Q1623-BX502 & 22.04 [22.69] & 0.3 & 203 & 15 & 24 & 79&  0.130 & 3.5 & 0.92 \\
Q1623-BX543 & 20.54 [20.66] & 0.4 & 10 & 431 & 189 & 98 &  0.285 & 4.6 & 0.92 \\
Q1700-BX490 & 19.99 [20.26] & 0.4 & 10 & 441 & 154 & 160 &  0.280 & 5.6 & 0.93 \\
Q1700-BX710 & 20.18 [20.23] & 4.4 & 1278 & 34 & 15 & 59 &  0.195 & 2.5 & 0.36 \\
Q1700-BX763 & 20.94 [20.98] & 1.4 & 905 & 15 & 6 & 23 &  0.125 & 0.8 & 0.36 \\
DSF2237a-C2 & 20.53 [20.71] & 2.9 & 255 & 112 & 39  & ... & 0.175 & ... & ... \\
Q2343-BX418 & 21.88 [22.29] & 0.1 & 102 & 11 & 17 & 32 &  0.030 & 1.4 & 0.93 \\
Q2343-BX513 & 20.10 [20.16] & 7.8 & 3000 & 26 & 13 & 44 &  0.155 & 1.8 & 0.19  \\
Q2343-BX660 & 20.98 [21.21] & 0.8 & 1609 & 5 & 19 & 30 &  0.005 & 1.8 & 0.69 \\
\hline
\multicolumn{10}{c}{Non-Detections}\\
\hline
Q0100-BX210 & ... & 0.8 & 102 & 74 & ... & ... &  0.215  & ... & ... \\
HDF-BX1311 & 20.48 & 0.9 & 255 & 35 & ... & 48 &  0.110 & 2.3 & 0.72 \\
HDF-BX1439 & 19.72 & 4.6 & 2200 & 21 & ... & 48 &  0.175 & 3.7 & 0.45 \\
Q1623-BX455 & 21.56 & 0.8  & 45  & 175 & ... & 169 &  0.370 & 5.4 & 0.87 \\
Q1623-BX663\tablenotemark{j} & 19.92 & 13.2  & 2000  & 21 & ... & 50 &  0.135 & 2.7 & 0.17 \\
Q1700-BX563 & ... & 3.3  & 641  & 51 & ...  & ... & 0.220  & ... & ... \\
Q1700-BX691 & 20.68 & 6.0  & 2750 & 22 & ...  & 36 & 0.285 & 2.1& 0.26 \\
Q2206-BX102 & ... & 11.8  & 2750  & 43 & ... & ... &  0.335 & ... & ... \\
Q2343-BX389 & 20.18 & 7.2  & 2750  & 26 & ...  & 80 & 0.265 & 4.7 & 0.39 \\
Q2343-BX442 & 19.85 & 11.8  & 2750  & 43 & ... & 44 &  0.285 & 3.2 & 0.21 \\
Q2343-BX587 & 20.12 & 3.3 & 719 & 46 & ... & 32 &  0.175 & 2.1 & 0.39 \\
\enddata
\label{SED.table}
\tablenotetext{a}{Observed Vega magnitude, values in brackets are corrected for line emission (where observed).}
\tablenotetext{b}{Typical uncertainty $<\sigma_{M_{\ast}}/M_{\ast}> = 0.4$.}
\tablenotetext{c}{Typical uncertainty $<\sigma_{\rm Age}/\textrm{Age}> = 0.5$.}
\tablenotetext{d}{SFR derived from stellar population model.  Typical uncertainty $<\sigma_{\rm SFR}/\textrm{SFR}> = 0.6$.}
\tablenotetext{e}{Extinction-corrected nebular SFR.  Values for Q1623-BX543 extrapolated from \othree.}
\tablenotetext{f}{Nebular SFR estimate from long-slit observations by Erb et al. (2006b), corrected for extinction and slit losses.}
\tablenotetext{g}{Typical uncertainty $<\sigma_{E(B-V)}/E(B-V)> = 0.7$.}
\tablenotetext{h}{Gas masses taken from Erb et al. 2006c.}
\tablenotetext{i}{Gas fraction.}
\tablenotetext{j}{AGN}
\end{deluxetable}

\clearpage


\begin{deluxetable}{lccccccc}
\tablecolumns{8}
\tablewidth{0pc}
\tabletypesize{\scriptsize}
\tablecaption{Other Galaxy Samples}
\tablehead{
\colhead{} & \colhead{} & \colhead{} & \colhead{$v_{\rm shear}$\tablenotemark{c}} & \colhead{$\Delta v_{\rm (ISM - neb)}$\tablenotemark{d}} & \colhead{$M_{\ast}$}\tablenotemark{e} & \colhead{} & \colhead{} \\
\colhead{Galaxy} & \colhead{$z$\tablenotemark{a}} & \colhead{$K_s$\tablenotemark{b}} & \colhead{(km s$^{-1}$)} & \colhead{(km s$^{-1}$)} & \colhead{($10^{10} M_{\odot}$)} & \colhead{$\mu$\tablenotemark{f}} & \colhead{$12 +$ log($O/H$)\tablenotemark{g}}
}
\startdata
\multicolumn{8}{c}{F{\"o}rster Schreiber et al. (2006)}\\
\hline
Q1623-BX376 & 2.4087 & 20.84 & 30 & $-62$ & 0.5 & 0.81 & $8.43\pm0.17$\\
Q1623-BX455 & 2.4071 & 21.56 & 55 & $-9$ & 0.8 & 0.87 & $8.41\pm0.12$\\
Q1623-BX502 & 2.1555 & 22.04 & 50 & $-48$ & 0.3 & 0.92 & $\lesssim 8.16$\\
Q1623-BX528 & 2.2684 & 19.75 & 65 & $-220$ & 12.2 & 0.18 & $8.62 \pm0.04$\\
Q1623-BX599 & 2.3313 & 19.93 & 50 & $-117$ & 6.7 & 0.37 & $8.55\pm0.08$\\
SSA22a-MD41 & 2.1710 & 20.5 & 140 & $189$ & 2.7\tablenotemark{h} & 0.43\tablenotemark{h} & $8.43\pm0.08$\\
Q2343-BX389 & 2.1728 & 20.18 & 235 & $-76$ & 7.0\tablenotemark{h} & 0.22\tablenotemark{h} & $8.54\pm0.08$\\
Q2343-BX610 & 2.2102 & 19.21 & 165 & $-112$ & 17\tablenotemark{h} & 0.15\tablenotemark{h} & $8.68\pm0.01$\\
Q2346-BX404 & 2.0298 & 20.05 & 20 & $-277$ & 40 & 0.31 & $8.50\pm0.07$\\
Q2346-BX405 & 2.0308 & 20.27 & 30 & $-79$ & 1.6 & 0.56 & $\lesssim 8.16$\\
Q2346-BX416 & 2.2406 & 20.30 & 70 & $37$ & 2.5 & 0.52 & $8.46\pm0.07$\\
Q2346-BX482 & 2.2569 & ... & 100 & $-175$ & 6.6\tablenotemark{h} & 0.31\tablenotemark{h} & $8.49\pm0.07$\\
\hline
\multicolumn{8}{c}{Genzel et al. (2006)}\\
\hline
BzK15504 & 2.3834 & 19.2 & 171 & ... & 7.7 & 0.36\tablenotemark{i} & ...\\
\hline
\multicolumn{8}{c}{Genzel et al. (2008)}\\
\hline
SSA22a-MD41 & 2.172 & 20.5 & 165 & $189$ & 2.7 & 0.43\tablenotemark{h} & $8.27\pm0.06$\\
Q2343-BX389 & 2.174 & 20.18 & 261 & $-76$ & 7 & 0.22\tablenotemark{h} & $8.51\pm0.04$\\
Q2343-BX610 & 2.211& 19.21 & 158 & $-112$ & 17 & 0.15\tablenotemark{h} & $8.66\pm0.02$\\
Q2346-BX482 & 2.258 & ... & 213 & $-175$ & 6.6 & 0.31\tablenotemark{h} & $8.35\pm0.06$\\
BzK6004 & 2.387 & 18.9 & 146 & ... & 58 & 0.05\tablenotemark{h} & $8.69\pm0.02$\\
\hline
\multicolumn{8}{c}{Stark et al. (2008)}\\
\hline
J2135-0102 & 3.07 & ... & 55 & ... & 0.6 & ... & 8.6\\
\enddata
\label{otherwork.table}
\tablenotetext{a}{Systemic redshift, values drawn from the indicated references.}
\tablenotetext{b}{Vega magnitude, uncorrected for line emission. Values based on our own observations except for BzK15504, BzK6004, SSA22a-MD41.}
\tablenotetext{c}{Values based on indicated references: $v_r$ from F{\"o}rster Schreiber et al. (2006), $v_c$ sin$i$ from Genzel et al. (2006) and Stark et al. (2008), $v_d$ sin$i$ from Genzel et al. (2008).}
\tablenotetext{d}{Kinematic offset between ISM absorption line redshift and nebular emission line redshift, calculated from our rest-UV spectra.}
\tablenotetext{e}{Stellar masses for F{\"o}rster Schreiber et al. (2006) galaxies from our broadband photometry (except where indicated), for all other samples
drawn from the indicated reference.}
\tablenotetext{f}{Combines tabulated $M_{\ast}$ with $M_{\rm gas}$ from D.K. Erb (priv. comm.) except where indicated.}
\tablenotetext{g}{Oxygen abundance drawn from the indicated reference.}
\tablenotetext{h}{Values from Genzel et al. (2008).}
\tablenotetext{i}{Values from Genzel et al. (2006).}
\end{deluxetable}

\end{document}